\documentclass{JHEP3} 
\JHEPspecialurl{http://jhep.sissa.it/JOURNAL/JHEP3.tar.gz}

\usepackage{epsfig,multicol,bbm,bm,amsmath}
\allowdisplaybreaks[2]

\title{B polarization of cosmic background radiation from 
second-order scattering sources}

\author{M. Beneke, C.~Fidler and K.~Klingm\"uller\\
Institut f\"ur Theoretische Teilchenphysik und Kosmologie\\ 
RWTH Aachen University\\
D - 52056 Aachen, Germany}

\preprint{Preprint TTK-11-01\\
February 4, 2011
}	

\abstract{
B-mode polarization of the cosmic background radiation is 
induced from purely scalar primordial sources at second order in 
perturbations of the homogeneous, isotropic universe. We calculate 
the B-mode angular power spectrum $C_l^{BB}$ sourced by the second-order 
scattering term in the full second-order Boltzmann equations 
for the polarized radiation phase-space density, which have recently become 
available. We find that at $l\approx 200$ the second-order effect 
is comparable to the first-order effect for a tensor-to-scalar 
ratio of $r=10^{-6}$, and to about $2\cdot 10^{-4}$ at $l\approx 1000$. 
It is always negligible relative to the weak-lensing 
induced contribution.
}

\dedicated{}

\begin{document}

\section{Introduction}

Polarization is expected to play a central role in future studies 
of the cosmic background radiation. The polarization patterns are 
usually split into a divergence-like E-mode and a curl-like B-mode 
\cite{Kamionkowski:1996zd,Seljak:1996gy}. 
B-mode polarization is a powerful diagnostic for primordial 
gravitational waves, i.e. tensor fluctuations of the metric, 
and thereby constrains inflation models directly. While E-mode 
polarization has already been detected  \cite{Kovac:2002fg,Bennett:2003bz} 
and is being observed with increasing precision 
\cite{Larson:2010gs,Komatsu:2010fb}, a B-mode signal remains elusive. 
This, together with precise information on the temperature 
anisotropy spectrum indicates some suppression of primordial 
tensor versus scalar perturbations, since B-modes are not 
generated by scalar perturbations.

The absence B-mode polarization when the primordial fluctuations are 
purely scalar holds, however, only in linear perturbation theory. 
If primordial tensor fluctuations are indeed suppressed, B-mode 
polarization generated from scalar sources in second order may 
constitute an important background to the search for primordial 
gravitational waves. While such an effect would naturally be 
expected to be relevant 
at tensor-to-scalar ratios of order $10^{-5}$, which is 
the size of perturbations in the microwave background, only a full 
second-order calculation can tell whether there are no 
enhancements. In this paper we compute a new second-order effect 
that contributes to B-mode polarization.

Several second-order sources of B-mode polarization have 
already been calculated in various approximations. The most important is 
the weak-lensing effect, reviewed in Ref.~\cite{Lewis:2006fu}, which converts 
E-mode polarization to B-mode polarization as the photons travel through 
the inhomogeneous universe~\cite{Zaldarriaga:1998ar}. 
Weak lensing becomes large at small scales, and at large values of 
the perturbation wave-vector $\bm{k}$ the perturbation series breaks
down. The usual treatment of weak lensing therefore avoids  
perturbation theory by considering the small deflection angles of the 
photon trajectories. Another effect that has been estimated is 
B-mode polarization from gravitational time delay \cite{Hu:2001yq} 
and from sources proportional to second-order vector and tensor metric
perturbations, which are themselves generated from the 
product of scalar perturbations~\cite{Mollerach:2003nq}. 

Recently the full second-order Boltzmann equations for the cosmological 
evolution of the polarized radiation distribution have become 
available~\cite{Beneke:2010eg,Pitrou:2008hy}, so that in principle 
it is possible to compute the power spectrum of B-mode polarization 
in second order generated from primordial scalar sources without 
approximations. While this is numerically challenging and beyond the 
scope of the present paper, we focus here on the novel sources of 
B-mode polarization that appear in the second-order collision term, 
which have not been estimated before. As in previous investigations 
of the weak-lensing and second-order metric perturbation effect, 
we calculate the amount of B-mode polarization from second-order 
scattering sources in isolation, that is, we set the sources of 
other effects to zero in the equations. We then compare the magnitude 
of the collisional effect to those previously known. Investigating 
the various effects in isolation may be viewed as a first step to 
solving the full equations, and is probably a good approximation, 
since the different effects are related to different periods in the 
evolution of the universe. In particular, the collision sources 
relevant to the present paper are active mainly during a short period 
around recombination, and later, during reionization. The reionization 
contribution will, however, not be considered here. 

The outline of the paper is as follows.
In Section \ref{sec:equations} we define the B-mode angular 
power spectrum when the radiation spectrum is not black-body. 
We briefly review the structure of 
the second-order polarized Boltzmann equations and discuss the terms 
that define the new collisional sources. A numerical method based 
on Green functions, which is suited to compute the two-point 
function that relates to the angular power spectrum $C_l^{BB}$ of 
B-mode temperature fluctuations, is derived in Section \ref{sec:method}. 
In Section \ref{sec:calc} we apply this method to calculate the 
power spectrum and discuss issues related to numerical checks and 
stability. Our results and conclusions are summarized in 
Section~\ref{sec:results}. Two appendices collect equations 
and derivations that supplement Sections~\ref{sec:equations} 
and \ref{sec:method}.

\section{Second-order scattering sources of 
CBR polarization} 
\label{sec:equations}

In this section we begin by defining the B-mode angular power spectrum 
and summarize the second-order Boltzmann equations that will be solved
subsequently.

\subsection{The B-mode angular power spectrum}

The cosmic background radiation (CBR) is described by the phase-space 
density matrix 
\begin{equation}
f_{ab}(\eta,\bm{x},\bm{q}) = \delta_{ab}\,f^{(0)}_I(q) + 
f_{ab}^{(1)}(\eta,\bm{x},\bm{q})+
f_{ab}^{(2)}(\eta,\bm{x},\bm{q})+\ldots,
\end{equation}
which we expand around the unpolarized, homogeneous black-body 
background distribution $f^{(0)}_I(q)$. Expressed in terms 
of co-moving phase-space momenta 
$\bm{q} = q\bm{n}$, 
$f_I^{(0)}(q)$ is time-independent and the black-body 
temperature $T_0$ is the temperature of the CBR today. The indices 
$a,b=\pm$ refer to the circular polarization basis and can 
be exchanged for the Stokes parameters $X=I,V,E,B$ by a linear 
transformation \cite{Beneke:2010eg}. We define a matrix of 
fractional temperature perturbations $\Theta_{ab}(\eta,\bm{x},\bm{q})$ 
through 
\begin{equation}
f_{ab}(\eta,\bm{x},\bm{q}) = \left[\exp\left(\frac{q}
{T_0 \,(1+\Theta(\eta,\bm{x},\bm{q}))}\right)-1\right]^{-1}_{ab},
\end{equation}
where the ``one'' in square parenthesis and in the argument of the 
exponential function must be 
interpreted as the 2 x 2 unit matrix, and we expand 
$\Theta(\eta,\bm{x},\bm{q}) =\Theta^{(1)}(\eta,\bm{x},\bm{n})+
\Theta^{(2)}(\eta,\bm{x},\bm{q})+\ldots$. The term ``temperature 
perturbation'' is slightly misleading, since in general 
$f_{ab}(\eta,\bm{x},\bm{q})$ is not a Bose-Einstein distribution 
when perturbations are included, and hence $\Theta(\eta,\bm{x},\bm{q})$ 
is not independent of the photon energy $q$. However, as is 
well-known, the first-order perturbation $\Theta^{(1)}(\eta,\bm{x},\bm{n})$ 
{\em is} independent of $q$, and the spectrum remains black-body 
at first order with a position- and direction-dependent temperature. 
At second order, however, distortions of the black-body spectrum 
are expected.

We wish to consider observables independent of $q$, which may be 
interpreted as temperature perturbations. To this end we 
define the radiation energy density normalized to the unperturbed 
energy density,
\begin{equation}
\label{eq:Deltadef}
\Delta_{ab}(\eta,\bm{x},\bm{n}) = 
\frac{\int dq q^3 f_{ab}(\eta,\bm{x},q\bm{n})}
{\int dq q^3 f_I^{(0)}(q)}.
\end{equation}
If the photons were unpolarized and assumed a black-body spectrum with 
temperature $T(\eta,\bm{x},\bm{n})$, then $\Delta(\eta,\bm{x},\bm{n})$ 
would be related to 
the temperature perturbation by 
\begin{equation}
\Delta(\eta,\bm{x},\bm{n}) = 
\left[\frac{T(\eta,\bm{x},\bm{n})}{T_0}\right]^4 
=(1+\Theta(\eta,\bm{x},\bm{n}))^4.
\end{equation}
In general we obtain the matrix identities 
\begin{eqnarray}
\Delta^{(1)}(\eta,\bm{x},\bm{n}) &=& 4 \Theta^{(1)}(\eta,\bm{x},\bm{n}),
\nonumber\\
\Delta^{(2)}(\eta,\bm{x},\bm{n}) &=& 
4 \overline{\Theta}^{(2)}(\eta,\bm{x},\bm{n}) 
+ 6 \left[\Theta^{(1)}(\eta,\bm{x},\bm{n})\right]^2,
\end{eqnarray}
among the perturbation coefficients, where
\begin{equation}
\overline{\Theta}^{(2)}(\eta,\bm{x},\bm{n}) 
= \frac{\int dq q^3 f_I^{(0)}(q) \left[\Theta^{(2)}(\eta,\bm{x},q\bm{n}) 
+\frac{1}{4} q\frac{\partial}{\partial q}
\Theta^{(2)}(\eta,\bm{x},q\bm{n}) \right]} 
{\int dq q^3 f_I^{(0)}(q)}.
\end{equation}
The quantity $\overline{\Theta}^{(2)}$ obviously equals $\Theta^{(2)}$
when the latter is $q$ independent. Otherwise it represents the 
temperature perturbation of a black-body distribution with the 
same energy density as the actual radiation distribution 
$f(\eta,\bm{x},\bm{q})$.

We define the multipole expansion coefficients of the fractional 
temperature perturbation here and today as usual by 
\begin{equation}
a_{lm} = a^{(1)}_{lm}+a^{(2)}_{lm}+\ldots 
= \int d\Omega(\bm{n})\,Y^{s*}_{lm}(\bm{n})\,\left[
\Theta^{(1)}(\eta_0,\bm{x}_0,\bm{n})+ 
\overline{\Theta}^{(2)}(\eta_0,\bm{x}_0,\bm{n}) +\ldots\right],
\quad
\end{equation}
noting the presence of the equivalent black-body temperature 
perturbation and the 2 x 2 matrix nature of $a_{lm}$. The spin 
of the spin-weighted spherical harmonic, $Y^{s}_{lm}(\bm{n})$, 
is $s=0$ for the diagonal $++$, $--$ components, $s=+2$ for $ab=+-$, 
and $s=-2$ for $-+$. In the circular 
polarization basis B-mode 
polarization is given by $i/2$ times the difference of 
the $+-$ and $-+$ components of the 2 x 2 density matrix, hence 
we define 
\begin{equation}
a_{B,lm}=\frac{i}{2}\,(a_{+-,\,lm}-a_{-+,\,lm}).
\end{equation}
The absence of B-mode polarization in first order implies that 
the $+-$ and $-+$ components of $f^{(1)}_{ab}$ and $\Theta^{(1)}_{ab}$
as well as  $[\Theta^{(1)}]^2_{ab}$ are equal, so $a^{(1)}_{B,lm} = 0$.
At second order
\begin{eqnarray}
a^{(2)}_{B,lm} &=& \frac{1}{4} 
 \int d\Omega(\bm{n})\,\frac{i}{2}\left[Y^{+2*}_{lm}(\bm{n})\,
\Delta^{(2)}_{+-}(\eta_0,\bm{x}_0,\bm{n}) -Y^{-2*}_{lm}(\bm{n})\,
\Delta^{(2)}_{-+}(\eta_0,\bm{x}_0,\bm{n})
\right]
\nonumber\\
&=&  \frac{1}{4} \int\frac{d^3\bm{k}}{(2\pi)^3}\,
e^{i\bm{k}\cdot \bm{x}_0}\,(-i)^l\sqrt{\frac{4\pi}{2l+1}}\,
\Delta^{(2)}_{B,lm}(\eta_0,\bm{k}),
\label{eq:a2lm}
\end{eqnarray}
where in the last line we introduced the Fourier modes and 
multipole coefficients of $\Delta^{(2)}_{B}(\eta_0,\bm{x}_0,\bm{n})$.

Our aim is to compute the B-mode angular power spectrum $C_l^{BB}$ given by 
the statistical average
\begin{equation}
\langle a_{B,lm} a^*_{B,l'm'} \rangle =
\delta_{l l'}\delta_{m m'}\,C_l^{BB}
\end{equation}
when the perturbations of the FRW background at first order are purely
scalar. Using Eq.~(\ref{eq:a2lm}) this average can be expressed in 
terms of 
\begin{equation}
\langle \Delta^{(2)}_{B,lm}(\eta_0,\bm{k})
\Delta^{(2)*}_{B,l'm'}(\eta_0,\bm{k}^\prime)\rangle 
= (2\pi)^3\delta^{(3)}(\bm{k}-\bm{k}^\prime)\,
P^{\Delta_B}_{ll',mm'}(k,\bm{\hat k})
\label{eq:defPDeltaB}
\end{equation}
in the form
\begin{equation}
\langle a_{B,lm} a^*_{B,l'm'} \rangle = 
\frac{1}{16}\,(-i)^l\,i^{l'} \,
\frac{4\pi}{\sqrt{(2l+1)(2l'+1)}} \int\frac{d^3\bm{k}}{(2\pi)^3} \,
P^{\Delta_B}_{ll',mm'}(k,\bm{\hat k}).
\label{eq:aageneral}
\end{equation}
The power spectrum of $\Delta_B$ 
defined here depends on the direction 
$\bm{\hat k}$ of the mode vector, since it refers to the fixed 
coordinate system of an observer and helicity $m$ defined with 
respect to the three-axis of this system. When the helicity
axis is chosen 
to be $\bm{\hat k}$ we obtain the simpler expression
\begin{equation}
\langle \Delta^{(2)}_{B,lm}(\eta_0,\bm{k})
\Delta^{(2)*}_{B,l'm'}(\eta_0,\bm{k}^\prime)\rangle_{|\bm{\hat k}\;\rm axis} 
= \delta_{m m'}\,(2\pi)^3\delta^{(3)}(\bm{k}-\bm{k}^\prime)\,
P^{\Delta_B}_{ll',m}(k)
\label{eq:defPDeltaBkhat}
\end{equation}
where the form of the right hand side of the equation follows from 
statistical isotropy and homogeneity and explicitly from the 
results of Section~\ref{sec:method}. In particular, the power spectrum 
$P^{\Delta B}_{ll',m}(k)$ depends only on the magnitude $k=|\bm{k}|$. 
It is this power spectrum that will be computed later on for $l'=l$. 
The two power spectra are related through the transformation 
(\ref{eq:Tlmrotation}) of the B-mode multipoles under rotations, 
which gives
\begin{equation}
P^{\Delta_B}_{ll',mm'}(k,\bm{\hat k}) = \frac{4\pi}{\sqrt{(2l+1)
    (2l'+1)}} \sum_{\tilde m} Y^{-\tilde m}_{lm}(\bm{\hat k}) 
Y^{-\tilde m *}_{l'm'}(\bm{\hat k}) \,P^{\Delta_B}_{ll',\tilde m}(k).
\end{equation}
Plugging this into Eq.~(\ref{eq:aageneral}) and using the 
orthogonality of the spin-weighted spherical harmonics, we 
obtain the familiar result
\begin{equation}
C_l^{BB} = \frac{1}{16} \frac{2}{\pi} 
\sum_{m=\pm 1,\pm 2} \,\int_0^\infty dk\,k^2 \,
\frac{P^{\Delta_B}_{ll,m}(k)}{(2l+1)^2}.
\label{eq:CBl}
\end{equation}
The sum is restricted to values $|m|\leq 2$, 
since the equations for $\Delta_{X,lm}^{(2)}(\bm{k})$ in the frame 
where $\bm{\hat k}$ coincides with the helicity axis do 
not contain source terms when $|m|>2$. 
The sum does not include $m=0$ since there are no scalar mode
contributions to B-mode polarization.

\subsection{Second-order equations}

In Ref.~\cite{Beneke:2010eg} we derived the second-order Boltzmann
equations for the photon phase-space densities
$f_{X,lm}(\eta,\bm{k},q)$. We now explain the structure of these 
equations and discuss the approximations we apply in this paper 
to isolate the second-order scattering sources. We are interested 
in the energy integrated distribution functions defined by
\begin{equation}
\label{eq:Deltadeflm}
\Delta^{(n)}_{X,lm}(\eta,\bm{k}) = 
\frac{\int dq q^3 f^{(n)}_{X,lm}(\eta, \bm{k},q)}
{\int dq q^3 f_I^{(0)}(q)},
\end{equation}
in particular in $\Delta^{(2)}_{B,lm}(\eta,\bm{k})$. 
The equation for second-order B-mode polarization  
when there are no first-order vector and tensor perturbations, hence
$\Delta^{(1)}_{B,lm}(\eta,\bm{k})=0$, is given by
\begin{eqnarray}
\frac{\partial}{\partial \eta}\Delta_{B,lm}^{(2)}(\bm{k}) &+& 
\sum_{\pm} \,(\mp i) \Delta_{B,(l \pm 1)m_1}^{(2)}(\bm{k})\,k^{[m_2]}
D_{m_1 m}^{\pm,l} + 
i \Delta_{E,l m_1}^{(2)}(\bm{k}) \,k^{[m_2]} D^{0,l}_{m_1 m}  
\nonumber \\
&+&i k_1^{[m_2]} \left(A^{(1)}-D^{(1)} \right)\!(\bm{k}_1)\,
\Delta_{E,lm_1}^{(1)}(\bm{k}_2) K_{m_1 m}^{0, l}  
\nonumber \\
&+&\,\bigg( ik_2^{[m_2]}\left(A^{(1)}-D^{(1)}\right)\!(\bm{k}_1)
+4ik_1^{[m_2]} A^{(1)}(\bm{k}_1) \bigg)
\,\Delta_{E,lm_1}^{(1)}(\bm{k}_2)D_{m_1 m}^{0,l} 
\nonumber \\
=|\dot{\kappa}|\,\bigg\{ &-& \Delta_{B,lm}^{(2)}(\bm{k}) 
+ v^{(1)}_{e,[m_2]}(\bm{k}_1)\Delta_{E,lm_1}^{(1)}(\bm{k}_2) D^{0,l}_{m_1 m} 
\nonumber \\
&-& \delta_{l2} \,\frac{\sqrt{6}}{5} v^{(1)}_{e,[m_2]}(\bm{k}_1)
\left(\Delta_{I,2m_1}^{(1)}-\sqrt{6}\Delta_{E,2m_1}^{(1)}\right)\!(\bm{k}_2) 
D^{0,2}_{m_1 m} \bigg\}.
\label{eq:Bpol}
\end{eqnarray}
Some comments on notation are in order. The equations above and below 
are given in conformal Newtonian gauge for phase-space densities defined 
in an inertial frame locally at rest and aligned 
with respect to the general coordinate 
system. $A^{(1)}$ and $D^{(1)}$ denote
the first-order scalar metric perturbations, $v_{e,[m]}$ the
velocity field of the baryon-electron fluid, and $|\dot\kappa|$ is 
proportional to the collision rate of Thomson scattering. The coupling
coefficients $D^{0,l}_{m_1 m}$ etc. are summarized in 
Eq.~(\ref{eq:couplingcoeffs}) and Ref.~\cite{Beneke:2010eg}. Products 
of first-order perturbations involve two different mode vectors 
 $\bm{k}_1$ and $\bm{k}_2= \bm k-\bm{k}_1$ and it is understood that 
a convolution integral over $\bm{k}_1$ is performed. Similarly,  
for given $m$ we imply $m_2= m-m_1$ and a sum over $m_1$. Indices 
in square brackets refer to helicity rather than 
Cartesian vector components with $k^{[0]} = i k_3$. We refer 
to Ref.~\cite{Beneke:2010eg} for further notational details.  

Eq.~(\ref{eq:Bpol}) and the corresponding equations for the radiation 
intensity and E-mode polarization contain a number of different
effects:
\begin{itemize}
\item In addition to the conformal time derivative the first line contains 
the effect of free streaming of radiation perturbations in the 
unperturbed background. This term converts vector and tensor E-mode 
perturbations into the corresponding B-mode perturbations 
and further excites higher multipoles from the $l=0,1,2$ ones 
after radiation ceases to be tightly coupled to the baryons. 
These terms must obviously be included in a calculation of the 
angular power spectrum observed today.
\item
The remaining two lines before the equality sign are the weak 
lensing and time delay contributions, which represent the 
effect of space-time inhomogeneity on the photon perturbations as 
they travel through the universe. 
This converts E-mode polarization into B-mode polarization 
even in the absence of vector and tensor perturbations and 
is known to generate significant B-mode polarization at large 
$l$~\cite{Zaldarriaga:1998ar}. We drop these known contributions, 
since we are interested in the effect of the 
collision source terms.
\item The equation for $\Delta_{I,lm}^{(2)}(\bm{k})$ contains 
second-order metric source terms involving products of 
first-order perturbations such as $A^{(1)}(\bm{k}_1)D^{(1)}(\bm{k}_2)$
and second-order perturbations $A^{(2)}(\bm{k})$, 
$\dot{B}^{(2)}_{[m]}(\bm{k})$ and $\dot{E}^{(2)}_{[m]}(\bm{k})$. 
The amount of B-mode polarization generated from some of these 
terms has been estimated in Ref.~\cite{Mollerach:2003nq}. Specifically,  
the vector and tensor perturbations 
${B}^{(2)}_{[m]}(\bm{k})$ and ${E}^{(2)}_{[m]}(\bm{k})$ 
induced at second-order from purely scalar perturbations 
have been evaluated and inserted into the first-order equations 
for free streaming to obtain the amount of B-mode polarization today. 
Again, since we are interested in the effect of the new collision 
source terms, we drop these terms.
\item The right-hand side of Eq.~(\ref{eq:Bpol}) is the second-order 
collision term. The first term in curly brackets is a universal 
relaxation term for all multipoles that in the absence of source 
terms drives the phase-space distribution to its equilibrium form 
and hence damps polarization. This term is already present in 
first order. In addition we find new scattering sources of B-mode 
polarization from the coupling of first-order scalar intensity and 
E-mode polarization to perturbations in the electron-baryon bulk 
velocity. These terms and corresponding new terms in the equations 
for $\Delta_{I,lm}^{(2)}(\bm{k})$ and  $\Delta_{E,lm}^{(2)}(\bm{k})$ 
are kept in our numerical calculation. They constitute the 
scattering sources, whose effect is computed here for the first time. 
We thus include the complete second-order collision term with 
one exception. We neglect terms of the form 
$|\dot{\kappa}|\,[\delta x_e/x_e]^{(1)}\times$ first order
perturbations, which arise from the perturbed ionization history.
\end{itemize}
We thus solve the following system of second-order equations:
\begin{eqnarray}
\frac{\partial}{\partial \eta}\Delta_{I,lm}^{(2)}(\bm{k}) &+& 
\sum_{\pm} (\mp i) \Delta_{I,(l \pm 1)m_1}^{(2)}(\bm{k})
k^{[m_2]}C_{m_1 m}^{\pm,l}
\nonumber\\ 
=|\dot{\kappa}|\,\bigg\{&-& \Delta_{I,lm}^{(2)}(\bm{k}) 
+\delta_{l0}\Delta_{I,00}^{(2)}(\bm{k})  
+4\delta_{l1} v_{e,[m]}^{(2)}(\bm{k}) 
+ \delta_{l2}\frac{1}{10}\left(\Delta_{I,2m}^{(2)}(\bm{k}) - 
\sqrt{6}\Delta_{E,2m}^{(2)}(\bm{k}) \right)
\nonumber\\
&+& \left(A^{(1)}(\bm{k}_1)+\delta_b^{(1)}(\bm{k}_1)\right)
\bigg(-\Delta_{I,lm}^{(1)}(\bm{k}_2) 
+\delta_{l0}\Delta_{I,00}^{(1)}(\bm{k}_2)  
+4\delta_{l1} v_{e,[m]}^{(1)}(\bm{k}_2)  
\nonumber\\
&& \hspace*{0.0cm}
+\,\delta_{l2}\frac{1}{10}\left(\Delta_{I,2m}^{(1)} - 
\sqrt{6}\Delta_{E,2m}^{(1)} \right)\!(\bm{k}_2)\bigg) 
\nonumber\\[0.1cm]
&+&  \sum_{\pm}(\mp 1) v^{(1)}_{e,[m_2]}(\bm{k}_1)
\Delta_{I,(l \pm 1)m_1}^{(1)}(\bm{k}_2) C^{\pm,l}_{m_1 m}
 \nonumber\\
&+&\delta_{l0}\,v^{(1)}_{e,[m_2]}(\bm{k}_1)
\left(2\Delta_{I,1m_1}^{(1)}-4v^{(1)}_{e,[m_1]}\right)\!(\bm{k}_2)
C^{+,0}_{m_1 m} \nonumber\\[0.2cm]
&+& 3\delta_{l1}\,v^{(1)}_{e,[m_2]}(\bm{k}_1)
\Delta_{I,0m_1}^{(1)}(\bm{k}_2)C^{-,1}_{m_1 m}   
\nonumber\\[0.2cm] 
&+&\delta_{l2} \,v^{(1)}_{e,[m_2]}(\bm{k}_1)
\left(7 v^{(1)}_{e,[m_1]}(\bm{k}_2)
-\frac{1}{2}\Delta_{I,1m_1}^{(1)}(\bm{k}_2)\right) C^{-,2}_{m_1 m}  
\nonumber\\
&+&\frac{1}{2} \delta_{l3}\,v^{(1)}_{e,[m_2]}(\bm{k}_1)
\left(\Delta_{I,2m_1}^{(1)}-\sqrt{6}\Delta_{E,2m_1}^{(1)}\right)(\bm{k}_2)
C^{-,3}_{m_1  m} 
\,\bigg\}
\label{eq:2ndorderI}
\end{eqnarray}
\begin{eqnarray}
\frac{\partial}{\partial \eta}\Delta_{E,lm}^{(2)}(\bm{k}) &+&
\sum_{\pm} (\mp i) \Delta_{E,(l \pm
  1)m_1}^{(2)}(\bm{k})k^{[m_2]}D_{m_1 m}^{\pm,l} - i \Delta_{B,l
  m_1}^{(2)}(\bm{k}) k^{[m_2]} D^{0,l}_{m_1 m}  
\nonumber\\
=|\dot{\kappa}|\,\bigg\{ &-& \Delta_{E,lm}^{(2)}(\bm{k}) -\delta_{l2}
\frac{\sqrt{6}}{10} \left(\Delta_{I,2m}^{(2)}(\bm{k}) -
  \sqrt{6}\Delta_{E,2m}^{(2)}(\bm{k}) \right) 
\nonumber\\
&+&\left(A^{(1)}(\bm{k}_1)+\delta_b^{(1)}(\bm{k}_1)\right)
\bigg(-\Delta_{E,lm}^{(1)}(\bm{k}_2) -\delta_{l2} \frac{\sqrt{6}}{10}
  \left(\Delta_{I,2m}^{(1)} - \sqrt{6}\Delta_{E,2m}^{(1)}
  \right)\!(\bm{k}_2)\bigg) 
\nonumber\\
&+& \sum_{\pm}(\mp 1) v^{(1)}_{e,[m_2]}(\bm{k}_1) \Delta_{E,(l \pm
  1)m_1}^{(1)}(\bm{k}_2) D^{\pm,l}_{m_1 m} 
\nonumber\\
&+& \delta_{l2} \frac{\sqrt{6}}{2}v^{(1)}_{e,[m_2]}(\bm{k}_1)\left(
  \Delta_{I,1m_1}^{(1)}(\bm{k}_2) -2
  v^{(1)}_{e,[m_1]}(\bm{k}_2)\right) C_{m_1 m}^{-,2} 
\nonumber \\[0.2cm]
&-& \delta_{l3}\frac{\sqrt
  6}{2}v^{(1)}_{e,[m_2]}(\bm{k}_1)\left(\Delta_{I,2m_1}^{(1)}-
  \sqrt{6}\Delta_{E,2m_1}^{(1)}\right)\!(\bm{k}_2)D^{-,3}_{m_1 m} 
\,\bigg\}
\label{eq:2ndorderE}
\end{eqnarray}
\begin{eqnarray} 
\frac{\partial}{\partial \eta}\Delta_{B,lm}^{(2)}(\bm{k}) &+& 
\sum_{\pm} (\mp i) \Delta_{B,(l \pm 1)m_1}^{(2)}(\bm{k}) k^{[m_2]}
D^{\pm,l}_{m_1 m}  + i \Delta_{E,l m_1}^{(2)}(\bm{k}) k^{[m_2]}D_{m_1
  m}^{0,l} 
\nonumber\\
=|\dot{\kappa}|\,\bigg\{ &-& \Delta_{B,lm}^{(2)}(\bm{k}) + 
v^{(1)}_{e,[m_2]}(\bm{k}_1)\Delta_{E,lm_1}^{(1)}(\bm{k}_2)
D^{0,l}_{m_1 m} 
\nonumber \\
&-& \delta_{l2} \frac{\sqrt{6}}{5}v^{(1)}_{e,[m_2]}(\bm{k}_1)
\left(\Delta_{I,2m_1}^{(1)}-\sqrt{6}\Delta_{E,2m_1}^{(1)}\right)\!(\bm{k}_2)
D^{0,2}_{m_1 m} \,\bigg\}
\label{eq:2ndorderB}
\end{eqnarray}
\begin{eqnarray}
\left(\frac{\partial}{\partial \eta}+H_C\right)
v^{(2)}_{e,[m]}(\bm{k})
&=&\,-\frac{|\dot{\kappa}|}{4 R}\,
\bigg\{4 v_{e,[m]}^{(2)}(\bm{k})- \Delta_{I,1m}^{(2)}(\bm{k})
\nonumber\\[0.2cm]
&& \hspace*{-2.5cm}+\,A^{(1)}(\bm{k}_1)
\left(4 v^{(1)}_{e,[m]}(\bm{k}_2)-\Delta_{I,1m}^{(1)}(\bm{k}_2)\right)
- v^{(1)}_{e,[m_2]}(\bm{k}_1)\Delta_{I,2m_1}^{(1)}(\bm{k}_2) 
C^{+,1}_{m_1 m}\nonumber\\[0.2cm] 
&& \hspace*{-2.5cm}
+ \,4 v^{(1)}_{e,[m]}(\bm{k}_1)
\Delta_{I,00}^{(1)}(\bm{k}_2) \bigg\}.
\label{eq:2ndorderv}
\end{eqnarray}
with $R=3 \bar\rho_b/(4 \bar\rho_\gamma)$. 
The equation for $v^{(2)}_{e,[m]}(\bm{k})$ must be included, since
this quantity appears in the collision term of the intensity
perturbation equation (\ref{eq:2ndorderI}), and we applied the 
same approximations to this equation as for the other three. 
That is, we neglect the second-order metric perturbation 
$B^{(2)}_{[m]}$, and for consistency also products of first-order 
metric perturbations (since $B^{(2)}_{[m]}$ is itself sourced 
by a product of such terms). The complete equation for $v^{(2)}_{e,[m]}$ 
and the baryon density perturbation $\delta_b^{(2)}=
\left[\delta \rho_b/\bar \rho_b\right]^{(2)}$ and the precise definition 
of these quantities is provided in Appendix~\ref{ap:equations}. 
Eqs.~(\ref{eq:2ndorderI}) to (\ref{eq:2ndorderv}) form a 
closed system together with the equations for the first-order
perturbations also summarized in the Appendix, which we solve 
in the following. We note that the approximations made are systematic 
in the sense that in the absence of collision terms 
($|\dot{\kappa}|\to 0$) all second-order perturbations vanish; 
hence we exclusively focus on collisional effects as intended.


\section{Solving the second-order equations} 
\label{sec:method}

For convenience, we introduce a compact notation summarizing the photon
equations Eq.~\eqref{eq:2ndorderI} to Eq.~\eqref{eq:2ndorderB}:
\begin{equation}
\label{toy}
\dot{\Delta}_{n}^{(2)} + k C_{nm}\Delta_{m}^{(2)}
= -|\dot{\kappa}|(\Delta_{n}^{(2)} -\varsigma_{nm}\Delta_{m}^{(2)}-S_{n}).
\end{equation}
The indices are multi-indices, $n = (X_n, l_n m_n)$, where $X = I, E, B,
v_e$ distinguishes between the photon modes and the baryon
velocity; $l$, $m$ are the multipole indices.  Repeated multi-indices are
summed over $X$, $l$, and $m$; $\bm{k}$ has been aligned with the
three-direction, $\bm{k}=k \bm{e}_3$. The dependence on $k$ of
$\Delta_n$, $S_n$, and the Green functions $G_{nm}$ introduced below 
will not be made explicit. Note that Eq.~\eqref{toy} does not
represent the electron equation
Eq.~\eqref{eq:2ndorderv}, i.e., it is only valid for $n \neq
(v_e, 1 m)$.

The matrix $C_{nm}$ describes the free-streaming coupling of each photon
multipole moment to its neighbours with $l \pm 1$, which leads to a gradual
excitation of the initially small large-$l$ moments.  The conformal time it 
takes for an excitation to propagate from multipole moment $l$ to 
moment $l \pm \Delta l$ is $\eta \approx \Delta l / k$. Free streaming 
also accounts for the conversion between E- and B-mode polarization for 
vector and tensor modes, which is the only possible source for
B-mode polarization at first order.

The first part of the scattering term, $-|\dot\kappa| \Delta_n^{(2)}$, is
responsible for the tight-coupling suppression. At early times the scattering
rate $|\dot\kappa|$ is large, so that a non-vanishing moment
$\Delta_n^{(2)}$ induces a large gradient driving it to zero as long as 
the first part of the scattering term is not cancelled by the 
remaining term $\varsigma_{nm}\Delta_{m}^{(2)}+S_n$.  
Since the coupling $\varsigma_{nm}$ vanishes for high multipole moments, 
only the monopole and dipole are not suppressed by the tight coupling 
of photons and baryons. At second order the quadrupole $\Delta_{I,2m}$ 
is also present due to a cancellation with 
$S_n$~\cite{Beneke:2010eg,Pitrou:2008hy,Bartolo:2006fj}, 
but there is no polarization in tight coupling~\cite{Beneke:2010eg}. 

The source $S_n$ contains the convolutions of first-order perturbations.  Other
than in the first-order equations, where there are only source terms for low
multipole moments, in the second order source $S_n$ there are contributions for
all multipoles.  Moreover, there is a source term for B-mode polarization
which is induced by first-order quadrupoles, including the large intensity
quadrupole, cf. Eq.~\eqref{eq:2ndorderB},
\begin{equation}
S_{B,lm}
=
\cdots -\delta_{l2} \frac{\sqrt{6}}{5}v^{(1)}_{e,[m_2]}(\bm{k}_1)
\Delta_{I,2m_1}^{(1)}(\bm{k}_2) D^{0,2}_{m_1 m} \cdots .
\end{equation}
Such a direct source term for B polarization does not exist at first order,
where B-mode polarization is only generated from free streaming of vector 
and tensor E-mode polarization.

\subsection{Line-of-sight integration}
At first order the line-of-sight integration is used to great 
success~\cite{Seljak:1996is,Hu:1997hp}. It is
the solution of
\begin{equation}
\dot{\Delta}_{n} + k C_{nm}\Delta_{m} = -|\dot{\kappa}|\Delta_{n} 
+\rho_n,
\end{equation}
where $C_{nm}$ are the free-streaming coefficients introduced in
Eq.~\eqref{toy} and $\rho_n$ includes all source terms. The solution to this
equation is given by
\begin{equation}
\label{lossolution}
\Delta_{n}(\eta_0)
= \int_0^{\eta_0}  d\eta \,e^{-\kappa(\eta)} 
j_{nm}(k(\eta_0-\eta))\rho_m(\eta),
\end{equation}
where the functions $j_{nm}$ are combinations of spherical Bessel 
functions and Clebsch-Gordan coefficients,
\begin{eqnarray}
\label{losgreen}
\nonumber
j_{nm}(x) &=&
\sum_{l_1} i^{l_n-l_1-l_m} \frac{(2l_n+1)(2l_1+1)}{2l_m+1}j_{l_1}(x)
H_{X_n X_m}^*(l_n-l_1-l_m)\\
&&\times
\left(\begin{array}{ccc}l_n & l_1 &l_m\\ m_n & 0 & m_m\end{array}\right)
\left(\begin{array}{ccc}l_n & l_1 &l_m\\ F_{X_n} & 0 & F_{X_m}\end{array}
\right).
\end{eqnarray}
The matrix $H_{XX'}$ is responsible for the mixing between $E$ and $B$
polarization in free streaming and is given by $H_{XX'}(\text{even})=
\delta_{XX'}$ and $H_{XX'}(\text{odd})=\delta_{XI}\delta_{X'I} + i
\delta_{XE}\delta_{X'B} -i\delta_{XB}\delta_{X'E}$~\cite{Beneke:2010eg}; 
$F_X=0$ when $X=I$ and $F_X=-2$ when $X=E,B$.  Clebsch-Gordan coefficients are
denoted by big round brackets. A derivation of this solution is given in
Appendix \ref{sec:Los}.  The functions $j_{nm}$ are always real since the
product $i^{l_n - l_1 - l_m} H_{X_n X_m}^*$ can be imaginary only for 
odd $l_n - l_1 - l_m$ and $X_n = X_m = I$, but in that case the second 
Clebsch-Gordan coefficient vanishes.

To employ this solution at second order we only need to use the appropriate
source terms. In Eq.~\eqref{toy} we identify
$\rho_n = |\dot\kappa|(S_n +\varsigma_{nm} \Delta^{(2)}_m)$
and we obtain the integral solution
\begin{eqnarray}\label{eq:los}
\Delta_{n}^{(2)}(\eta_0)
&=&
\int_{0}^{\eta_0}d\eta\,|\dot{\kappa}(\eta)|e^{-\kappa(\eta)}
j_{nm}(k(\eta_0-\eta)) (S_{m}+\varsigma_{mp}\Delta_{p}^{(2)})(\eta).
\end{eqnarray}
Despite the second-order term in the integrand, this is a big step towards the
computation of second-order quantities today: the coefficient $\varsigma_{mp}$
vanishes for multipole moments $l_p > 2$ so that we are left with the task of
finding solutions for $\Delta_{p}^{(2)}(\eta)$ with $l \leq 2$.  In addition,
the visibility function $|\dot{\kappa}|e^{-\kappa}$ is non-zero only around
recombination and thus it is sufficient to compute these solutions for early
times $\eta \lesssim 500\,$Mpc/$c$. (We quote conformal time in units of
Mpc/$c$, but set the speed of light $c=1$ in equations.)


\subsection{Solution using Green functions}
The Boltzmann equations have to be solved with stochastic initial
conditions resulting from inflation.  At first order, the linearity of
the equations allows to write the solution as product of transfer
functions and the primordial fluctuations. The problem reduces to the task of
computing the non-stochastic transfer functions. This straightforward
separation is no longer possible at second order due to the quadratic
convolution terms. However, we can achieve a similar separation by using
Green functions.

The second-order photon equation Eq.~\eqref{toy} is linear in the 
second-order quantities,
\begin{eqnarray}
\dot{\Delta}^{(2)}_n(\eta) = A_{nm}(\eta)\Delta^{(2)}_m(\eta) + \sigma_n(\eta),
\end{eqnarray}
where $A_{nm} = |\dot\kappa| (\varsigma_{nm} - \delta_{nm}) - k C_{nm}$ and
$\sigma_n = |\dot\kappa| S_n$. The same applies to the electron-velocity
equation with $A_{nm}$ and $\sigma_n$ according to
Eq.~\eqref{eq:2ndorderv}.  The solution of such a linear differential equation
can be written in terms of Green functions,
\begin{align}
\Delta_n^{(2)}(\eta) =&
G_{nm}(\eta,\eta_\mathrm{ini}) \Delta_m^{(2)}(\eta_\mathrm{ini})
+ \int_{\eta_\mathrm{ini}}^\eta d\eta' \,
G_{nm}(\eta,\eta') \sigma_m(\eta'),
\label{eq:greenansatz}
\end{align}
where the  Green function ${\cal G}_{nm}(\eta,\eta') = 
G_{nm}(\eta,\eta')\theta(\eta-\eta')$ satisfies
\begin{equation}
\label{eq:greeneq}
\partial_\eta {\cal G}_{nm}(\eta,\eta')
= A_{np}(\eta){\cal G}_{pm}(\eta,\eta') + \delta_{nm}\delta(\eta-\eta').
\end{equation}
In this differential equation there are no stochastic quantities, thus we
can compute the Green function $G_{nm}(\eta,\eta')$ by solving the equation 
\begin{equation}
\label{eq:greeneqshort}
\partial_\eta G_{nm}(\eta,\eta') = A_{np}(\eta)G_{pm}(\eta,\eta')
\end{equation}
for $\eta>\eta'$ 
with initial condition $G_{nm}(\eta',\eta') = \delta_{nm}$ numerically using
standard methods. 
Note that with $\bm{k}=k\bm{e}_3$ aligned 
with the three direction and $i k^{[0]} = -k$, the matrix 
$A_{np}(\eta)$ is real and therefore the Green functions are real as
well.

In principle, Green functions could be used to calculate all two-point
functions for any $\eta$ until today, 
but generating the Green functions for late times and large
multipoles numerically is very time-consuming. A much better performance can be
achieved by combining this ansatz with the line-of-sight integration.  Using
the Green function method we compute 
$\Delta_p^{(2)}(\eta)$ for $l \leq 2$ and for early times, and 
then use Eq.~\eqref{eq:los} (with lower limit $\eta=0$ replaced by 
$\eta_\mathrm{ini}$, since the visibility function is negligibly small at 
very early times) to obtain the late-time
evolution of the higher multipole moments:
\begin{eqnarray}
\Delta_{n}^{(2)}(\eta_0) &=&
\int_{\eta_\mathrm{ini}}^{\eta_0}  
d\eta \,|\dot{\kappa}(\eta)| e^{-\kappa(\eta)}
j_{nm}(k(\eta_0-\eta))
\,\Big[
\varsigma_{mp}G_{pq}(\eta,\eta_\mathrm{ini}) \Delta_q^{(2)}(\eta_\mathrm{ini})
\nonumber\\[0.2cm]
&&+
\int_{\eta_\mathrm{ini}}^{\eta}d\eta'\left(
\delta_{mq}\delta(\eta-\eta')+|\dot\kappa(\eta')|\,
\varsigma_{mp}G_{pq}(\eta,\eta')\right)
S_q(\eta')
\Big].
\label{eq:finalsolution}
\end{eqnarray}
Note that besides $\Delta_q^{(2)}(\eta_\mathrm{ini})$ only the source
$S_q(\eta')$ depends on the stochastic initial conditions whereas the evolution
is given by the non-stochastic functions $j_{nm}$ and $G_{pq}$.
This separation is convenient for computing second-order correlation 
functions as described in the following subsections.

\subsection{Second-order power spectra}

The result Eq.~\eqref{eq:finalsolution} simplifies further if the initial
second-order quantities $\Delta_q^{(2)}(\eta_\mathrm{ini})$ vanish.  As we will
discuss in section \ref{sec:initial}, this is the case for the computation of
the non-scalar modes with $m = \pm1, \pm2$ and only these are needed in
Eq.~\eqref{eq:CBl} to compute the B-mode angular power-spectrum. Consequently,
we evaluate the two-point function assuming $\Delta_q^{(2)}(\eta_\mathrm{ini})
= 0$ and obtain
\begin{eqnarray}
&&\langle
\Delta_{n}^{(2)}(\bm k,\eta_0) \Delta_{n'}^{(2)*}(\bm k',\eta_0)
\rangle
\nonumber\\[0.2cm]
&&\hspace*{1cm}
=\,\int_{\eta_\mathrm{ini}}^{\eta_0}d\eta 
\int_{\eta_\mathrm{ini}}^{\eta_0}d\eta'\,
|\dot{\kappa}(\eta)||\dot{\kappa}(\eta')|\,e^{-\kappa(\eta)-\kappa(\eta')}
j_{nm}(k(\eta_0-\eta))j_{n'm'}(k(\eta_0-\eta'))
\nonumber\\[0.1cm] 
&& \hspace*{1.5cm}
\times \,
\int_{\eta_\mathrm{ini}}^{\eta}d\eta_1\int_{\eta_\mathrm{ini}}^{\eta'}d\eta_1'
\left(
\delta_{mq}\delta(\eta-\eta_1)
+ |\dot\kappa(\eta_1)|\varsigma_{mp}G_{pq}(\eta,\eta_1)
\right)
\nonumber\\[0.3cm]
&& \hspace*{1.5cm}
\times \,\Big(
\delta_{m'q'}\delta(\eta'-\eta_1')
+ |\dot\kappa(\eta_1')|\varsigma_{m'p'}G_{p'q'}(\eta',\eta_1')
\Big)
\langle S_{q}(\bm k,\eta_1)S_{q'}^*(\bm k',\eta_1')\rangle,
\qquad
\label{eq:2pfunc}
\end{eqnarray}
where we anticipated Eq.~(\ref{Finaleq}) below that sets
$\bm{k}=\bm{k}'$ in the integrand. We also imply that 
the helicity axis is the direction of $\bm{k}$ or, equivalently, 
that $\bm{k}$ is aligned with the three-axis of the 
coordinate system. Comparison with 
Eq.~(\ref{eq:defPDeltaBkhat}) 
provides the power spectrum of $\Delta_B^{(2)}$ required to 
compute the B-mode angular power spectrum (\ref{eq:CBl}).
The source term correlation function 
$\langle S_q(\bm k,\eta_1)S_{q'}^*(\bm k',\eta_1')\rangle$ can be
calculated from first-order quantities only. These are given by the primordial
potential $\Phi(\bm{k})\equiv A^{(1)}(\bm{k},\eta_{\rm ini})$,
multiplied by a transfer function $T$. 
Using $\bm{k}_2 = \bm{k}-\bm{k}_1$ we can write
\begin{eqnarray}
S_n(\bm{k},\eta)
&=&
\int \frac{d^3k_1}{(2\pi)^3} \,K_n^{pq}
\Delta_p^{(1)}(\bm{k}_1,\eta)\Delta_q^{(1)}(\bm{k}_2,\eta)
\nonumber\\
&=&
\int \frac{d^3k_1}{(2\pi)^3} \,K_n^{pq}\,
T_p^{(1)}(\bm{k}_1,\eta) T_q^{(1)}(\bm{k}_2,\eta)\Phi(\bm{k}_1)\Phi(\bm{k}_2),
\label{eq:source}
\end{eqnarray}
where the constants $K_n^{pq}$ are determined by the expressions for the 
second-order collision term, Eqs.~(\ref{eq:2ndorderI}) to (\ref{eq:2ndorderv}).
Thus, to compute the expectation value on the right-hand side of
Eq.~\eqref{eq:2pfunc} we need to evaluate a four-point function of the
primordial potential 
\begin{eqnarray}
	&&\langle
	\Phi(\bm k_1) \Phi(\bm k_2) \Phi^*(\bm k_1') \Phi^*(\bm k_2')
	\rangle
	\nonumber\\[0.2cm]&&\quad=
	\langle \Phi(\bm k_1) \Phi(\bm k_2) \rangle
	\langle \Phi^*(\bm k_1') \Phi^*(\bm k_2') \rangle
	+
	\langle \Phi(\bm k_1) \Phi^*(\bm k_1') \rangle
	\langle \Phi(\bm k_2) \Phi^*(\bm k_2') \rangle
	\nonumber\\[0.1cm]&&\qquad+\,
	\langle \Phi(\bm k_1) \Phi^*(\bm k_2') \rangle
	\langle \Phi(\bm k_2) \Phi^*(\bm k_1') \rangle
	\nonumber \\[0.2cm]&&\quad=
	(2 \pi)^6
	\Big(
	P_\Phi(k_1) P_\Phi(k_1')
	\delta^{(3)}(\bm k)
	\delta^{(3)}(\bm k')
	\nonumber\\&&\qquad+
	P_\Phi(k_1) P_\Phi(k_2)
	\Big[
	\delta^{(3)}(\bm k_1 - \bm k_1')
	\delta^{(3)}(\bm k_2 - \bm k_2')
	+
	\delta^{(3)}(\bm k_1 - \bm k_2')
	\delta^{(3)}(\bm k_2 - \bm k_1')
	\Big]
	\Big),\qquad
\label{eq:wick}
\end{eqnarray}
where $P_\Phi$ is the primordial power spectrum,
$ \langle \Phi(\bm k) \Phi^*(\bm k') \rangle = 
(2 \pi)^3 \delta^{(3)}(\bm k - \bm k') P_\Phi(k)$, and 
Gaussian statistics of the primordial potential $\Phi(\bm k)$ 
was assumed.
For $\bm k \neq 0$ or $\bm k' \neq 0$ this results in
\begin{equation}
\langle
\Phi(\bm k_1) \Phi(\bm k_2) \Phi^*(\bm k_1') \Phi^*(\bm k_2')
\rangle
= (2 \pi)^6 P_\Phi(k_1) P_\Phi(k_2)
\delta^{(3)}(\bm k - \bm k')
\Big[
\delta^{(3)}(\bm k_1 - \bm k_1')
+\delta^{(3)}(\bm k_1 - \bm k_2')
\Big],
\end{equation}
and the complete expectation value on the right-hand side of
Eq.~\eqref{eq:2pfunc} reads
\begin{eqnarray}
\langle S_n(\bm{k},\eta) S^*_{n'}(\bm{k}\,',\eta')\rangle
&=&
(2\pi)^3\delta^{(3)}(\bm{k}-\bm{k}\,') \int \frac{d^3k_1}{(2\pi)^3}
P_{\Phi}(k_1)P_\Phi(k_2)K_n^{pq} K_{n'}^{p'q'*} \nonumber\\
&&\times \bigg[ T_p^{(1)}(\bm{k}_1,\eta)T_q^{(1)}(\bm{k}_2,\eta)
T_{p'}^{(1)*}(\bm{k}_1,\eta') T_{q'}^{(1)*}(\bm{k}_2,\eta') 
\nonumber\\[0.1cm]
&& \hspace*{0.3cm}
+ \,T_p^{(1)}(\bm{k}_1,\eta)T_q^{(1)}(\bm{k}_2,\eta)
T_{p'}^{(1)*}(\bm{k}_2,\eta')T_{q'}^{(1)*}(\bm{k}_1,\eta') \bigg].
\qquad
\label{Finaleq}
\end{eqnarray}

For a given primordial spectrum, we can now compute the second-order power
spectrum using Eq.~\eqref{eq:2pfunc} and Eq.~\eqref{Finaleq}.

\subsection{Non-Gaussianity}

Analogously, the method described above can be applied to study 
non-Gaussianity. At first order, the bispectrum (three-point function) 
is always proportional to the primordial bispectrum and
therefore zero if the primordial perturbations are Gaussian.  This is 
different, if non-linear effects are taken into account. The leading 
contributions to the bispectrum are terms combining one second-order 
perturbation with two first-order perturbations,
\begin{eqnarray}
\langle\Delta_n \Delta_m \Delta_p \rangle
&=&
\langle\Delta_n^{(2)} \Delta_m^{(1)} \Delta_p^{(1)} \rangle
+ \text{sym}.
\end{eqnarray}
We replace the first-order quantities by their transfer functions and the
primordial potential and calculate the second-order quantity using the
combination of line-of-sight integration and Green functions,
Eq.~\eqref{eq:finalsolution}. Allowing for non-vanishing initial 
values $\Delta_n^{(2)}(\bm k, \eta_\mathrm{ini})$ we find, for 
$\bm{k}$ aligned with the three-axis of the coordinate system, 
\begin{eqnarray}
&& \langle \Delta_n^{(2)}(\bm k,\eta_0)\Delta_q^{(1)}(\bm k',\eta_0)
\Delta_r^{(1)}(\bm k'',\eta_0)\rangle
=
T_q^{(1)}(\bm k',\eta_0)T_r^{(1)}(\bm k'',\eta_0)
\nonumber\\[0.2cm]
&& \hspace*{0.6cm}\times \int_{\eta_\mathrm{ini}}^{\eta_0}  
d\eta\,|\dot{\kappa}| e^{-\kappa}j_{nm}(k(\eta_0-\eta))
\,\bigg[
	\varsigma_{mp}G_{pq}(\eta,\eta_\mathrm{ini})
	\langle
	\Delta_q^{(2)}(\bm k, \eta_\mathrm{ini}) \Phi(\bm k')\Phi(\bm k'')
	\rangle
\nonumber\\[0.1cm]
&&  \hspace*{1cm}+\,\int_{\eta_\mathrm{ini}}^{\eta}d\eta'\left(
	\delta_{mq}\delta(\eta-\eta')+|\dot\kappa(\eta')|
	\varsigma_{mp}G_{pq}(\eta,\eta')
	\right)
	\langle
		S_q(\bm k,\eta')\Phi(\bm k')\Phi(\bm k'')
	\rangle
\bigg].\qquad
\end{eqnarray}
Using Eq.~\eqref{eq:source} and contracting the resulting four-point function
similar to Eq.~\eqref{eq:wick} we can write the second expectation value as
\begin{eqnarray}
\langle S_n(\bm k,\eta')\phi(\bm k')\phi(\bm k'')\rangle
&=& (2\pi)^3\delta^{(3)}
(\bm k + \bm k'+ \bm k'')P_\Phi(k')P_\Phi(k'')K_n^{pq} 
\nonumber\\[0.2cm] 
&& \hspace*{-2cm}\times\,
\Big[T_p^{(1)}(-\bm k',\eta')T_q^{(1)}(-\bm k'',\eta') + 
T_p^{(1)}(-\bm k'',\eta')T_q^{(1)}(-\bm k',\eta')\Big].
\qquad
\end{eqnarray}
For an initial value $\Delta_q^{(2)}(\bm k,\eta_\mathrm{ini})$ which is
quadratic in the primordial potential $\Phi$, also the first expectation value
can be contracted and written in terms of the primordial power-spectrum. 
A detailed study of non-Gaussianity from second-order effects employing 
these equations will be presented in a follow-up article.

\section{Numerical evaluation}
\label{sec:calc}
In this section we outline the steps required to compute the two-point
function~\eqref{eq:2pfunc} and the $C_l^{BB}$ angular power spectrum 
(\ref{eq:CBl}) numerically.  First we need to compute the 
first-order transfer functions, which is discussed in 
Section~\ref{sec:first}, and the Green functions, as discussed in 
Section~\ref{sec:green}. In Section~\ref{sec:source} we obtain the 
source terms and combine both results to perform the line-of-sight 
integral together with the wave-vector convolutions.

For all intermediate results presented in this section we use a $\Lambda$CDM
model with massless neutrinos and the 
following parameters: $T_\mathrm{CBR} = 2.726\,\mathrm K$, 
$H_0 = 70\,\mbox{km}/(\mbox{s}\,\mbox{Mpc})$, and 
$\Omega_\mathrm{CDM}= 0.245$,
$\Omega_\mathrm{baryon}= 0.045$, $\Omega_\Lambda = 0.71$. 
We assume a scale-invariant primordial power-spectrum
$P_\Phi(k) = 2 \pi^2 (9 / 25) \Delta_\mathcal R^2 / k^3$
with amplitude
$\Delta_\mathcal R^2 = 2.41 \times 10^{-9}$.
To compute the recombination history we employ {\sc Recfast}
\cite{Seager:1999bc} with a helium mass-fraction $Y_\mathrm p = 0.24$. Our
calculation does not include late effects such as the late integrated
Sachs-Wolfe effect (ISW) and reionization. Scattering effects and the early 
ISW will be considered up to $\eta=500 \,\mathrm{Mpc}/c$
and neglected for later times where we use the free-streaming approximation.
Recombination occurs around $\eta_{\rm rec} = 286.7 \,\mathrm{Mpc}/c$, defined
as the median of the visibility function.

\subsection{First-order solutions} \label{sec:first}
The first-order transfer functions can be computed by solving the first-order
Boltzmann equations with standard methods for solving ordinary differential
equations (e.g. as implemented in the GNU Scientific Library \cite{gsl}).
Alternatively one can resort to programs like {\sc Cmbfast}
\cite{Seljak:1996is}, {\sc Camb} \cite{Lewis:1999bs} or {\sc Cmbeasy}
\cite{Doran:2003sy}. The following results are based on our own 
code that has been compared to {\sc Camb}.

A crucial point is that we need only the low multipoles at early times where
the numerical calculation is straightforward. The restriction to early
times is due to the factor $|\dot\kappa|$ in all collision term
sources which is negligible after recombination. The reason why only 
low multipoles are needed at early times is that higher moments are
only slowly generated by free streaming. Thus, one can cut the
Boltzmann hierarchy at some $l_\mathrm{cut}$ and
neglect all higher multipole moments. Alternatively, to save CPU time,
one can apply the cut at a much lower multipole where the corresponding 
moment cannot be neglected, but is replaced by a closing relation. This 
option will be discussed in detail in Section~\ref{sec:close}.

Neutrinos start free streaming much earlier than photons because they are
not tightly coupled to baryons by Thomson scattering and thus start 
generating higher multipoles at early times. This has to be taken into 
account by choosing a cut
at larger $l$ for the neutrino hierarchy than for the photon
hierarchy. Typically we take more than 50 neutrino multipoles into account
which yields accurate results until matter domination.


\subsection{Initial conditions}
\label{sec:initial}

We assume that the primordial first-order perturbations are adiabatic and begin
the evolution of the transfer functions at $a_{\rm ini} = 10^{-6}$
(corresponding to $\eta_\mathrm{ini} = 0.464 \,\mathrm{Mpc} / c$)
deep in the radiation era with standard adiabatic initial conditions 
for the scalar perturbations:
\begin{eqnarray}
&&
T^{(1)}_A(k) \equiv T^{(1)}_\Phi(k) = 1,
\qquad
T^{(1)}_D(k)= - \left(1+\frac{2}{5}\frac{\bar \rho_\nu}
{\bar\rho_\gamma+\bar\rho_\nu}\right),
\nonumber\\
&&
T^{(1)}_{\delta_b}(k)=T^{(1)}_{\delta_c}(k) = 
\frac{3}{4} T^{(1)}_{\Delta_{I,00}}(k) = 
\frac{3}{4} T^{(1)}_{\Delta_{\nu,00}}(k) = 
-\frac{3}{2},
\nonumber\\[0.1cm]
&&
T^{(1)}_{v_{e,[0]}}(k) = T^{(1)}_{v_{c,[0]}}(k) 
=\frac{1}{4} T^{(1)}_{\Delta_{I,10}}(k) 
= \frac{1}{4} T^{(1)}_{\Delta_{\nu,10}}(k) 
= \frac{k}{2 H_C}.
\label{1stinitial}
\end{eqnarray}
For the neutrinos we further include an initial quadrupole 
$T^{(1)}_{\Delta_{\nu,20}}(k) = 2 k^2/(3 H_C^2)$, which  
however is very small at $\eta_{\rm ini}$, when for all $k$ of 
interest $k/H_C\ll 1$.

Setting the initial conditions for the second-order perturbation 
variables is simplified by the fact that by Eq.~(\ref{eq:CBl}) we 
only need to compute the vector and tensor perturbations 
$m=\pm 1, \pm 2$ when working in the frame where the mode-vector $\bm{k}$ is 
aligned with the helicity axis (usually the three-direction). 
While we may adopt the convention 
that at some initial time $A^{(2)}$ vanishes, since any non-zero 
value can be absorbed into a small change of  $A^{(1)}$, this 
cannot generally be done for all perturbation variables. For 
instance, deep in the radiation era, when $k/H_C\ll 1$, the total 
energy density perturbation is given by 
\begin{equation}
\frac{\delta\rho^{(2)}}{\bar \rho} = -2 A^{(2)} +4 H_C 
[A^{(1)}]^2.
\end{equation}
However, the equations we need to solve for $m=\pm 1, \pm 2$ do 
not contain the second-order energy density perturbations, that 
is the monopoles of the multipole decomposition, and hence we do 
not have to determine their initial conditions.

We do need the $m=\pm 1$ components of the second-order electron velocity 
$v_{e,[m]}^{(2)}$ and the radiation intensity dipole. Their 
initial values are related to the metric vector perturbation $B^{(2)}_{[m]}$. 
Since, as discussed above, we do not consider the second-order 
metric perturbations in this paper, the initial value  
of the second-order velocities and dipoles is consistently set 
to zero. In the tightly coupled radiation era there exists a 
non-vanishing radiation quadrupole $\Delta^{(2)}_{I,2m}$, which 
acquires a non-zero initial condition given in terms of  the square of 
the first-order electron velocity. Being of second order and 
suppressed by $(k/H_C)^2$, this can be safely neglected. We show 
in Section~\ref{sec:testtight} that collisions quickly drive the 
quadrupole to its tight-coupling value when it is zero initially. Finally, 
we note that the second-order neutrino perturbations do not 
appear in our equations, so we do not have to set their initial 
values. 

To sum up, we may solve the second-order equations with all initial 
values of second-order variables set to zero.

\subsection{Green functions}
\label{sec:green}
The differential equations~\eqref{eq:greeneqshort} for the Green functions 
are far less complicated than the original second-order equations. 
In particular, the equations are no longer stochastic.
Their structure is identical to the one of the 
first-order Boltzmann equations. 

The relevant Green functions $G_{nm}(\eta,\eta')$ can be classified by
several criteria. All Green functions except those for monopole,
dipole, quadrupole and electron velocity  are strongly
tight-coupling suppressed.\footnote{This does not apply to the corresponding
calculation of second-order neutrino perturbations.} 
This leads to a suppression of these
Green functions if $\eta'$ is located before
recombination (except for $\eta$ close to $\eta'$), and allows us to
restrict the
$\eta'$ integration to a period directly around recombination, as after 
recombination the source terms vanish with $|\dot\kappa|$ and before 
recombination the Green function is small, even after multiplication 
with the large scattering rate. 
The effect resembles that of the visibility function
$|\dot\kappa|e^{-\kappa}$ in the line-of-sight integration which is peaked
around recombination. 

Green functions which are not suppressed during tight-coupling may lead to
numerical difficulties because they in turn do not suppress the source
$\sigma_n = |\dot\kappa| S_n$ in Eq.~\eqref{eq:greenansatz}. At early times, 
the scattering rate $|\dot\kappa|$ is huge and, multiplied with the numerical
error of $S_n$, yields a large absolute error of $\sigma_n$ and the
corresponding second-order quantities. As a consequence, linear combinations of
these second-order quantities can obtain a large relative error, if 
there exist cancellations. For the electron velocity and radiation 
dipole Green functions such problems can be avoided by
exploiting the close relation of dipole and electron-velocity sources. The
electron-velocity sources can be split into two parts, one which 
is -- apart from a
prefactor -- identical to the photon-intensity dipole source and a remaining
part $\tilde{S}_{v_e,1m}$ which cancels the factor of $|\dot{\kappa}|$ by the
tight-coupling suppression $1/|\dot{\kappa}|$ of the combination of first-order
perturbations multiplying it. Writing 
\begin{equation}
S_{v_e, 1m} = -\frac{1}{4R}\,S_{I,1m} +\tilde{S}_{v_e, 1m},
\label{eq:Svsplit}
\end{equation}
where 
\begin{equation}
\tilde{S}_{v_e, 1m} = \frac{1}{4R} \,\delta_b^{(1)}(\bm{k}_1)
\left(4 v_{e,[m]}^{(1)}(\bm{k}_2)-\Delta_{I,1m}^{(1)}(\bm{k}_2)\right),
\end{equation}
the first part on the
right-hand side of Eq.~(\ref{eq:Svsplit}) 
can be combined with the photon source in Eq.~(\ref{eq:greenansatz}) 
to obtain 
\begin{eqnarray} \nonumber
\Delta_n^{(2)}(\eta)
&=&
\int_{\eta_\mathrm{ini}}^\eta \,d\eta^\prime \,|\dot\kappa(\eta')|
\left(G_{n,(I,1m)}(\eta,\eta')S_{I,1m}(\eta') + 
G_{n,(v_e,1m)}(\eta,\eta')S_{v_e,1m}(\eta')
\right) \\ \nonumber
&&+\sum \limits_{\text{all other }p} 
\,\int_{\eta_\mathrm{ini}}^\eta \,d\eta' \,|\dot\kappa(\eta')|\,
G_{np}(\eta,\eta')S_{p}(\eta') \\[0.2cm] \nonumber
&=&
\int_{\eta_\mathrm{ini}}^\eta \,d\eta'\,|\dot{\kappa}(\eta')|\,
\left(
 G_{n,(I,1m)}(\eta,\eta')- \frac{1}{4 R} \,G_{n,(v_e, 1m)}(\eta,\eta')
\right)S_{I,1m}(\eta') \\ \nonumber
&&
+ \int_{\eta_\mathrm{ini}}^\eta \,d\eta' \,|\dot{\kappa}(\eta')| 
\,G_{n,(v_e,1m)}(\eta,\eta') \tilde S_{v_e,1m}(\eta') 
\nonumber\\
&& + \sum \limits_{\text{all other }p} \,
\int_{\eta_\mathrm{ini}}^\eta \,d\eta' \,|\dot\kappa(\eta')|\,
G_{np}(\eta,\eta')S_{p}(\eta').
\end{eqnarray}
Thus, only the combination 
$G_{n,(I,1m)}- \frac{1}{4R}G_{n,(v_e,1m)}$ is
multiplied with the source $S_{I,1m}$ which is large at early times, 
but this combination of Green functions does vanish in the 
tight-coupling regime.
In Fig.~\ref{CombinedGreen} the combination is plotted illustrating the
suppression during tight-coupling (solid line).

\FIGURE[t]{
\epsfig{file=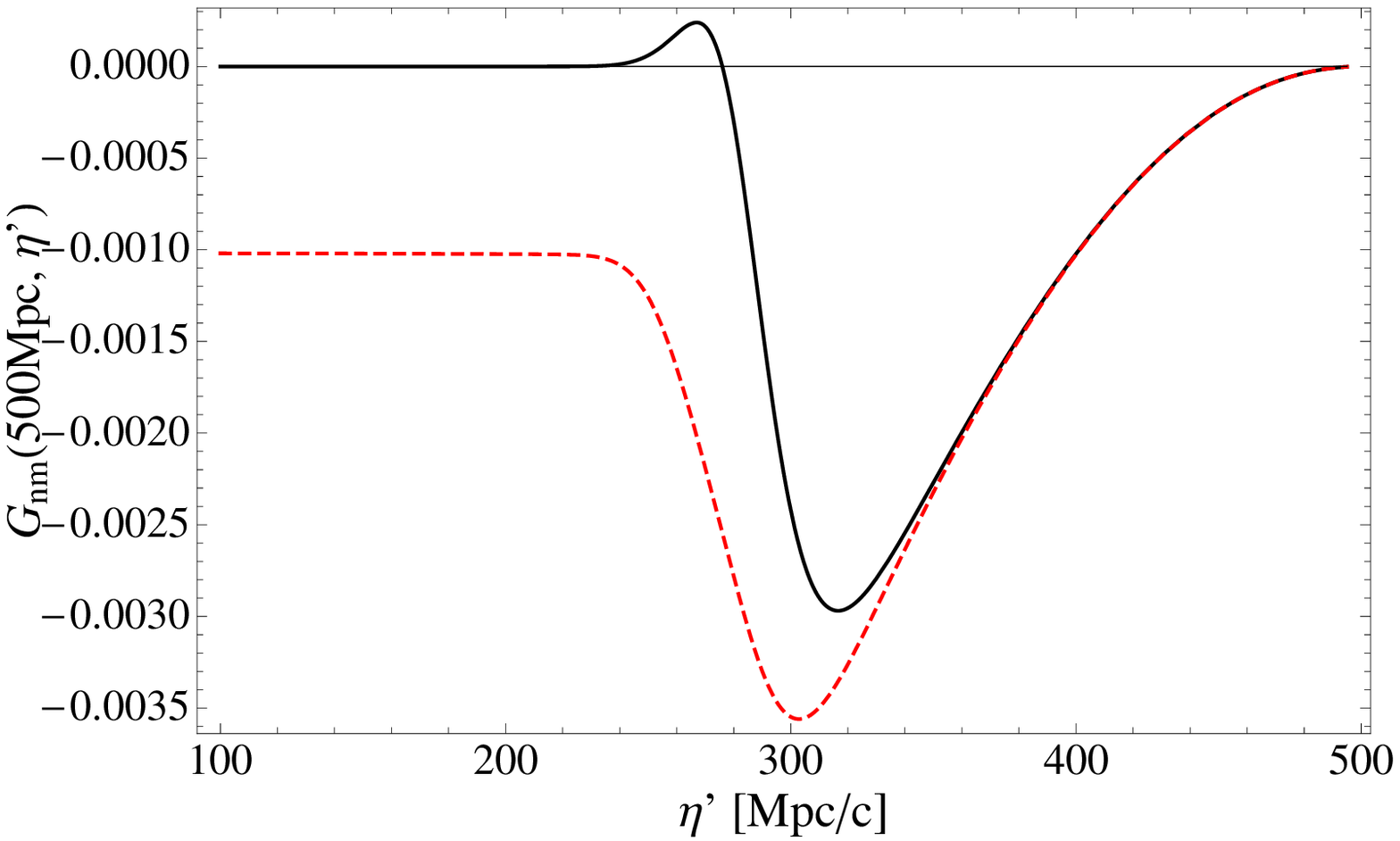,width=0.6\linewidth}
\caption{\label{CombinedGreen}
In dashed (red) the Green function $G_{(E,21),(I,11)}(500\,
\mathrm{Mpc}/c,\eta')$ for $\bm k = k \bm e_3$, $k=0.02 \,\mathrm{Mpc}^{-1}$.
Since the dipole is not tight-coupling suppressed, the Green function 
does not vanish before recombination. After recombination the Green 
function vanishes because $\Delta_{E,21}$ and $\Delta_{I,11}$ are only 
coupled in the scattering term. The solid (black) line is the combined 
dipole and electron-velocity Green function
$G_{(E,21),(I,11)}(500\,
\mathrm{Mpc}/c,\eta')-\frac{1}{4 R} G_{(E,21),(v_e,11)}(500\,
\mathrm{Mpc}/c,\eta')$ which vanishes at early times.}
}

The remaining Green functions which do not vanish in tight-coupling
after multiplication with  $|\dot{\kappa}|$ include: 
the Green function acting on the photon monopole, the one
acting on the remaining part of the electron-velocity source $\tilde{S}_{v_e,
1m}$ and the Green functions acting on the quadrupoles. The first does
not enter our computation since the photon monopole does
not couple to polarization, the second is suppressed by $\tilde{S}_{v_e, 1m}$
which itself is tight-coupling suppressed as explained above. Finally,
the Green functions acting on the quadrupoles are only suppressed by
one power of $|\dot{\kappa}|$ for $|m|=1$, and do not vanish when
multiplied by $|\dot{\kappa}|$.\footnote{The Green functions for $|m|=2$
are exponentially suppressed as there is no coupling to the
unsuppressed dipole.} Hence, while for $|m|=2$ the $\eta'$ integration in 
Eq.~\eqref{eq:greenansatz} can be restricted to times
around recombination as there are no contributions at early times, 
the time integration cannot be cut for $|m|=1$. This difference between
vector and tensor perturbations is related to the persistence of the 
dipoles during tight-coupling. While  during tight-coupling no
polarization is generated, the short-lived excitation of a quadrupole will 
modify the dipoles. This dipole is then converted
into a quadrupole during recombination which couples to
polarization. This indirect coupling of early vector sources can
induce polarization while early tensor sources 
have no influence, as there is no unsuppressed moment. 

After recombination, the characteristics of a Green function depend
on the coupling between the Green function's two multipoles.
If both moments are only coupled in the scattering
term, the Green function quickly vanishes after recombination.
If they are coupled via free streaming, it 
oscillates, since free streaming converts neighboring
moments into each other. In combination with the factor
$|\dot\kappa(\eta')|$ in Eq.~\eqref{eq:finalsolution} both types of Green
functions decay after recombination. However, the former decays
with an additional factor of $|\dot\kappa|$,
as depicted in Fig.~\ref{fig:greenf1}.

\FIGURE[t]{
\epsfig{file=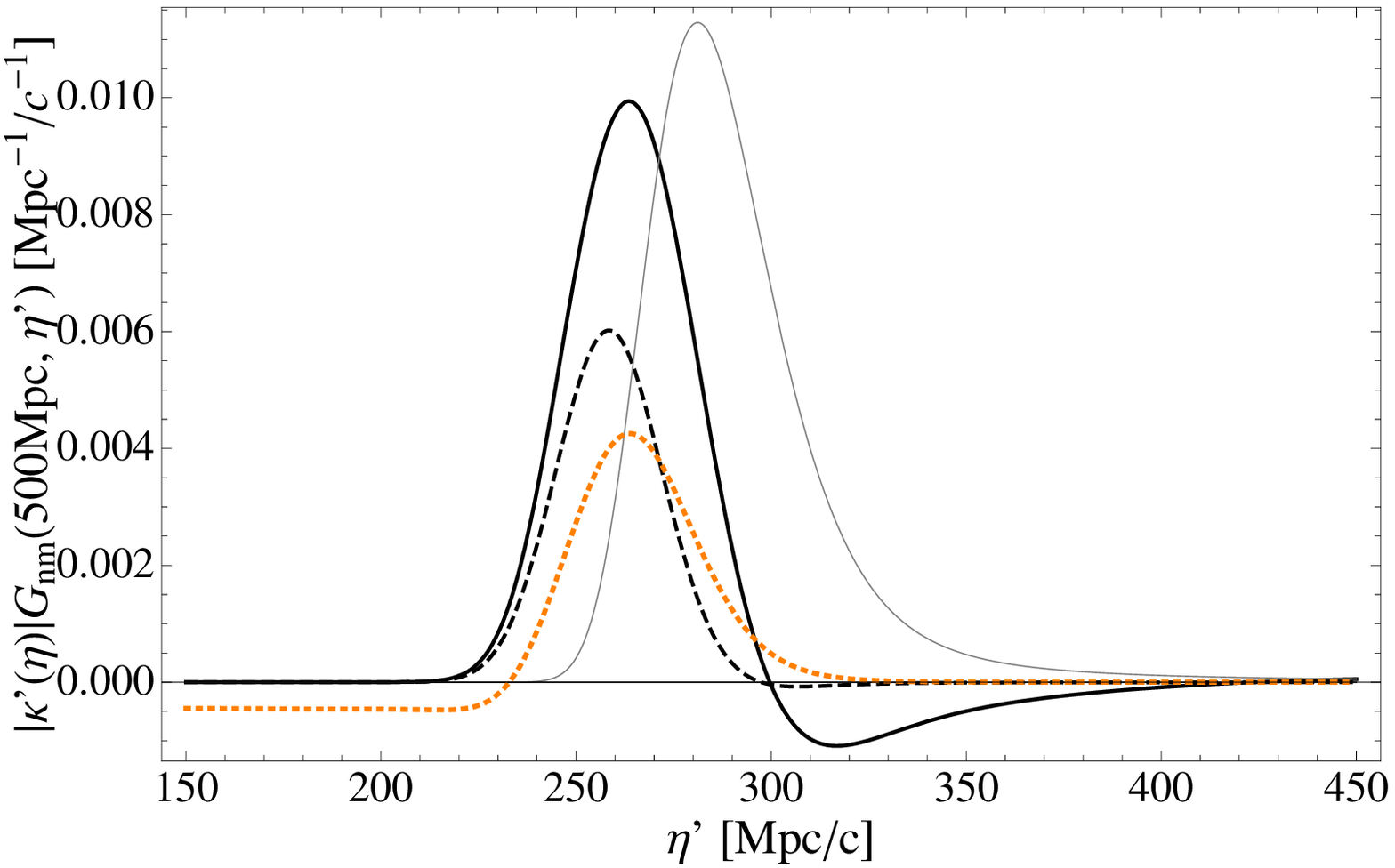,width=0.6\linewidth}
\caption{
Green functions $G_{(E,22),(E,22)}(500 \,\mathrm{Mpc}/c,\eta')$, $ 10\times G_{(I,22),(E,22)}(500
\,\mathrm{Mpc}/c,\eta')$ and $G_{(I,21),(E,21)}(500
\,\mathrm{Mpc}/c,\eta')$ (solid/black, dashed/black and dotted/orange, respectively) for
$\bm k = k \bm e_3$, $k=0.03 \,\mathrm{Mpc}^{-1}$, multiplied with 
$|\dot{\kappa}(\eta')|$. The multipole moments
$\Delta_{E,2m}$ and $\Delta_{I,2m}$ are only coupled in the 
scattering term. As expected, the
Green function connecting these moments vanishes faster than the 
Green function connecting moments coupled in the free-streaming term. 
The Green function acting on quadrupoles for $|m|=1$ is only
suppressed by $|\dot{\kappa}|$ and thus the plotted combination is 
constant at early times. The visibility function divided by two is
shown in grey for comparison.
\label{fig:greenf1}}
}

\subsection{Closing relations}
\label{sec:close}
Like the first-order Boltzmann hierarchy, the equations for the Green
functions need to be truncated at some multipole moment $l_\mathrm{cut}$.  The
straightforward approach is to set higher moments with $l > l_\mathrm{cut}$ to
zero. This is a good approximation at early times because all higher moments
are initially zero. But once the moment $l_\mathrm{cut}$ is excited by free
streaming, the solutions for the Green functions and first-order quantities,
respectively, become inaccurate, since in the full hierarchy a non-zero
moment at $l_\mathrm{cut}$ excites the moment $l_\mathrm{cut} + 1$ which
then feeds back into the lower moments.  However, the error will first 
become noticeable only in the moments close to $l_\mathrm{cut}$. It then 
takes an equal amount of time to carry the error back
to the lowest multipoles as it has taken to excite $l_\mathrm{cut}$ in the
first place. Hence, one can still trust the results for the lowest multipoles
some time after the highest multipole has been excited.

For accurate results one has to cut at some sufficiently large 
$l_\mathrm{cut}$ such that the lowest moments (which we are interested in) 
are not influenced by the truncation. To compute the photon multipole $l$ 
at time $\eta$ and wave number $k$, the rule of thumb is to 
cut at $l_\mathrm{cut} \approx l + k \,(\eta-\eta_\mathrm{rec}) / 2$. Thus,
if $k$ or $\eta$ is large, many multipoles have to be taken into account. This
can be avoided by using appropriate closing relations at a comparably 
small~$l$~\cite{Seljak:1996is}. 

Approximating the visibility function by a delta function at
$\eta_\mathrm{rec}$ allows us to derive an analytic relation between the
highest multipoles we consider. In this approximation the scattering rate is
infinitely large before recombination so that all multipoles except
for the 
monopole and dipole are zero due to tight-coupling suppression.  After
recombination, the scattering rate and therefore the sources are zero and we
can employ the free-streaming solution Eq.~\eqref{eq:los1} to obtain
\begin{eqnarray}\label{freestream} \nonumber
\Delta_{X,lm}(\eta) & = & \sum_{l_2=0,1} 
\sum_{X^\prime,l_1} i^{l - l_1 - l_2}
\,\frac{(2l + 1)(2l_1 + 1)}{2l_2 + 1}\,j_{l_1}
(k(\eta - \eta_\mathrm{rec})) \\
&& \times\,\left(\begin{array}{ccc}l & l_1 &l_2 \\ m & 0 & m\end{array}\right)
\left(\begin{array}{ccc}l&l_1&l_2\\F_{X'}&0&F_{X'}\end{array}\right)
\,H_{XX'}^*(l - l_1 - l_2)\Delta_{X',l_2m}(\eta_\mathrm{rec}).
\end{eqnarray}
Starting from this equation we perform the following steps to derive the
closing relations, exemplified here for the intensity multipoles 
$\Delta_{I,l0}$ and $l_\mathrm{cut}=9$.\footnote{The $m = 0$ calculation is 
not required to compute B-mode polarization, but the expressions are less
complicated and the method can be extended straightforwardly to other values 
of $m$.}
\begin{itemize}
\item We write down Eq.~(\ref{freestream}) explicitly for the two highest
multipole moments below $l_\mathrm{cut}$,
$\Delta_{X,(l_\mathrm{cut}-1)m}$ and $\Delta_{X,(l_\mathrm{cut}-2)m}$:
\begin{eqnarray}
 \Delta_{I,80}(\eta)
 &=&
  \left (8 j_{7}- 9 j_9\right) \Delta_{I,10}(\eta_\mathrm{rec})
  +17  j_8 \Delta_{I,00}(\eta_\mathrm{rec}) \nonumber \\
 \Delta_{I,70}(\eta)
 &=&
  \left (7 j_{7}- 8 j_9\right)\Delta_{I,10}(\eta_\mathrm{rec})
  +15  j_8\Delta_{I,00}(\eta_\mathrm{rec})
\end{eqnarray}
For brevity we leave out the argument of the Bessel functions which 
is always $k(\eta-\eta_\mathrm{rec})$. For $m \neq 0$, the monopole 
(and even the dipole for $m=\pm 2$) does not exist, and one uses instead the 
two lowest non-vanishing multipoles. For polarization, 
one uses the E- and B-mode quadrupole.
\item We solve both equations to write the monopole and dipole at 
recombination in terms of the multipoles $\Delta_{X,(l_\mathrm{cut}-1)m}$ and
$\Delta_{X,(l_\mathrm{cut}-2)m}$. In case of polarization ($X=E,B$), 
we have to relate $\Delta_{E,(l_\mathrm{cut}-1)m}$ and 
$\Delta_{B,(l_\mathrm{cut}-1)m}$ to $\Delta_{E,2m}$ and $\Delta_{B,2m}$ 
instead, due to the mixing of $E$ 
modes and $B$ modes during free streaming. 
\item In Eq.~(\ref{freestream}) for
$\Delta_{X,l_\mathrm{cut}m}$ we can now replace the monopole and dipole
on the right-hand side with the result from the previous step to obtain 
the closing relation. In our example we find
\begin{eqnarray} \nonumber
\label{Eq:closed}
\Delta_{I,90}(\eta)
&=&
\frac{19j_9\left(-7j_6+8j_8\right)+15j_7\left(9j_8-10j_{10}\right)}
{17j_8\left(-7j_6+8j_8\right)+15j_7\left(8j_7-9j_9\right)}
\,\Delta_{I,80}(\eta) \\
&&
+\frac{19j_9\left(8j_7-9j_9\right)+17j_8\left(-9j_8+10j_{10}\right)}
{17j_8\left(-7j_6+8j_8\right)+15j_7\left(8j_7-9j_9\right)}
\,\Delta_{I,70}(\eta).
\end{eqnarray}
\item Finally, we use this result to replace $\Delta_{X,l_\mathrm{cut} m}$ in
the Boltzmann hierarchy and obtain a closed system of differential
equations for all moments up to $\Delta_{X,(l_\mathrm{cut}-1)m}$ which
is independent of higher moments.
\end{itemize}
The quality of this approximation depends on the width of the visibility
function. While the photon evolution is dominated by scattering effects,
neutrino modes are exclusively sourced by metric terms. Consequently, the
closing relations described here cannot be applied to neutrinos, since they
rely on the sharply peaked visibility function. 
However, we only have to consider neutrino perturbations at first order,
so that all Green functions can be computed using the closing relations.

In our calculation of the Green functions we cut the hierarchy 
at $l_\mathrm{cut}= 9$, obtaining the closing relation
\begin{equation}
\label{eq:closing1}
 \Delta_{I,9m}(\eta)
 = h_{I,8m}(k(\eta-\eta_\mathrm{rec}))\Delta_{I,8m}(\eta)
 + h_{I,7m}(k(\eta-\eta_\mathrm{rec})) \Delta_{I,7m}(\eta).
\end{equation}
In general the functions $h_{X,lm}$ are combinations of spherical Bessel
functions as in the $m = 0$ example, Eq.~(\ref{Eq:closed}).  
For $m \neq 0$ the functions become more complicated, but as long as the
argument $k(\eta-\eta_\mathrm{rec})$ is small, they can be approximated by
polynomials.  This is not possible for larger arguments where the functions
oscillate, see Fig.~\ref{fig:closing1}. Only for very large arguments the
oscillations can be neglected due to damping and a simple approximation can be
used again. 

\FIGURE[t]{
\epsfig{file=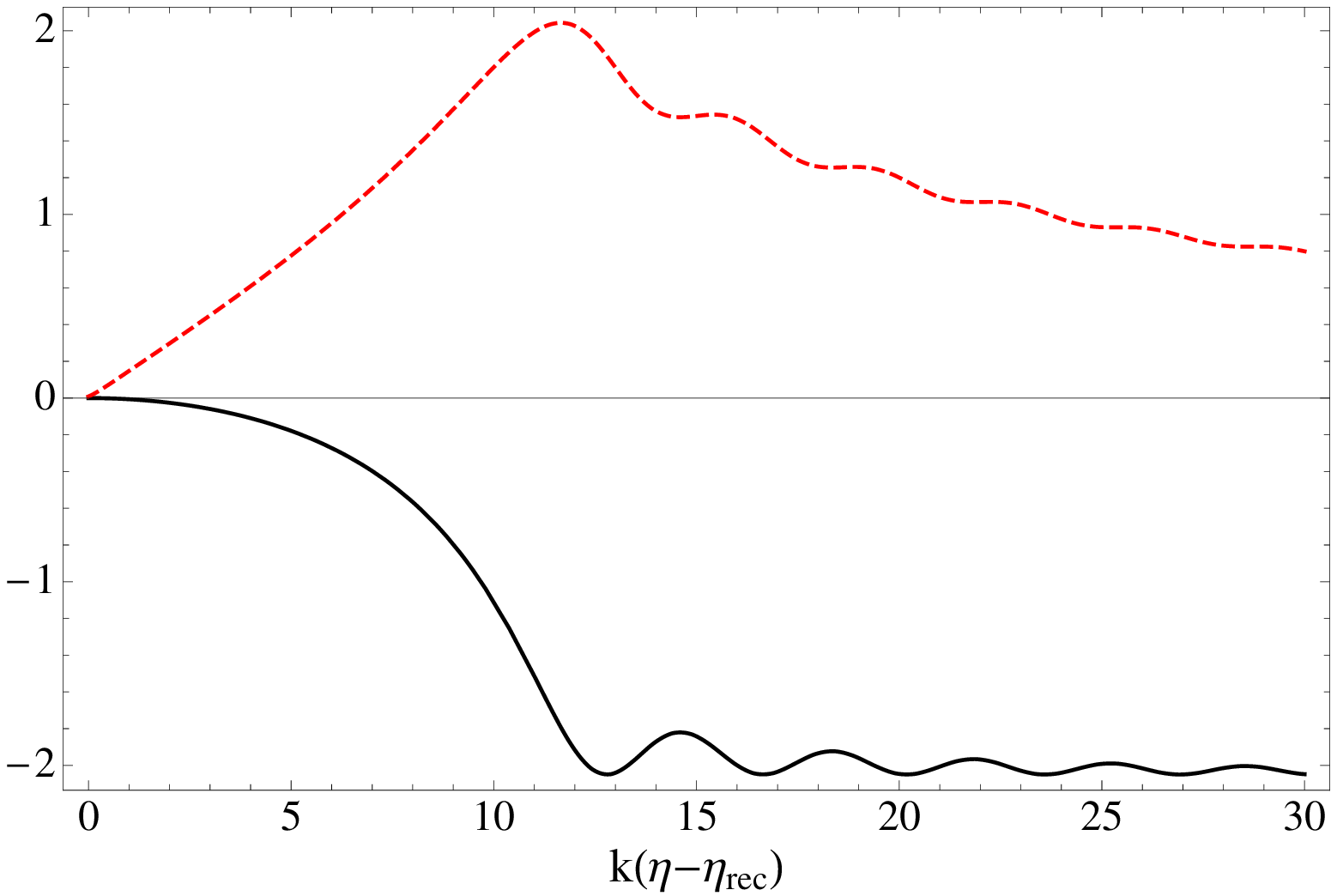,width=0.6\linewidth}
\caption{
\label{fig:closing1}
The functions $h_{I,72}$ (solid/black) and $h_{I,82}$ (dashed/red) 
which appear in the closing relation~\protect\eqref{eq:closing1}.}
}
\FIGURE[t]{
\epsfig{file=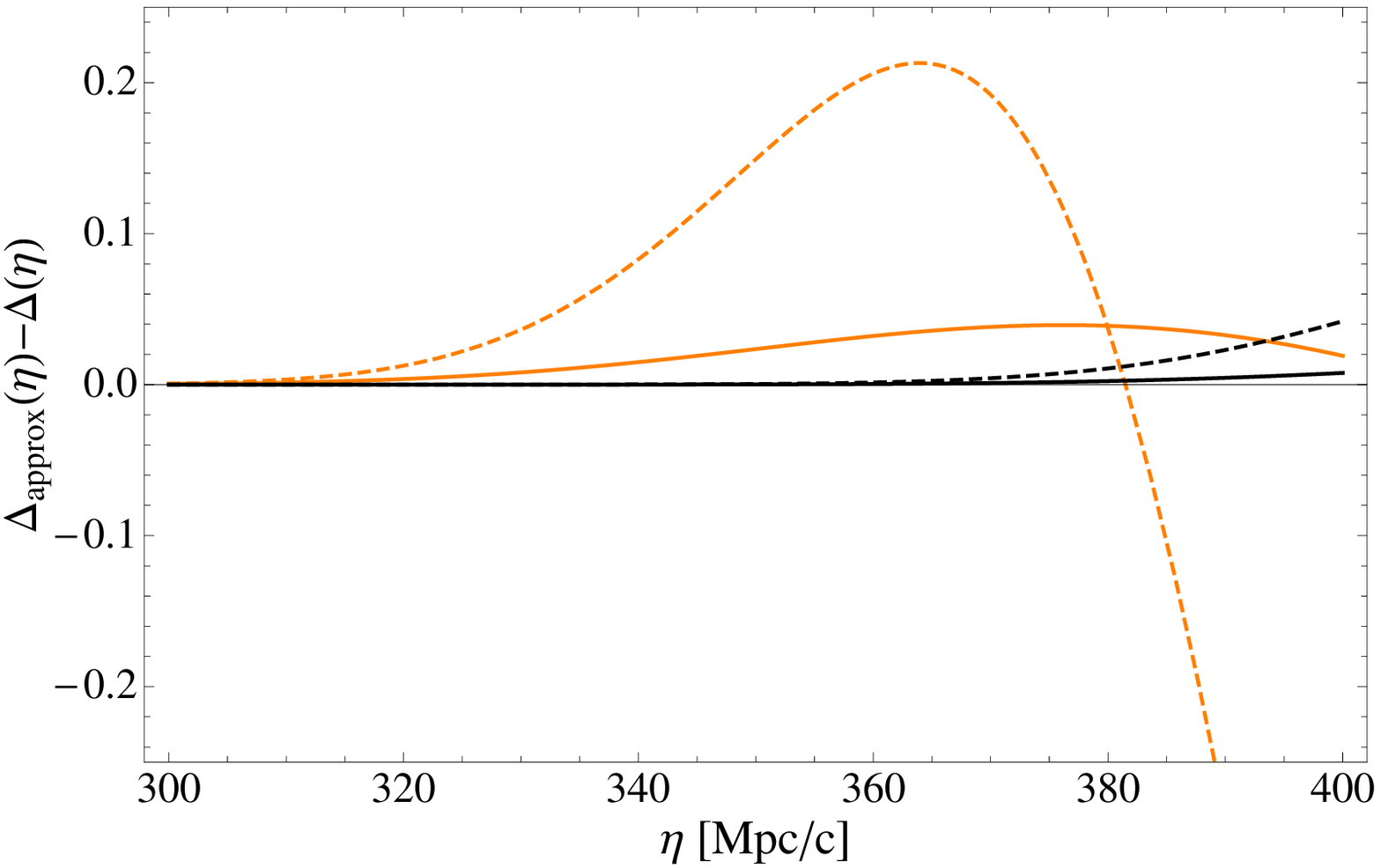,width=0.6\linewidth} 
\caption{\label{fig:closing}
Comparison of the absolute error obtained by using a simple truncation
(dashed) and the closing
relations (solid) for $\Delta_{I80}^{(1)}$ (black) and $\Delta_{I20}^{(1)}$
(orange); $k = 0.1 \,\mathrm{Mpc}^{-1}$, $l_\mathrm{cut} = 9$.
The improvement achieved by using closing relations is even better for
smaller wave numbers.
}
}

Figure \ref{fig:closing} shows that the closing relations can be used to
significantly reduce the error compared to a simple truncation --- at 
almost no additional CPU time.

\subsection{Source terms and integration}
\label{sec:source}
To compute the angular power spectrum $C_l$ at second order, 
we have to perform eight integrations:
two time integrals from the line-of-sight solution, another two time
integrals from the Green function ansatz, three integrations from the
convolution over $\bm k_1$ and finally the $k$ integral in Eq.~\eqref{eq:CBl}.
The integral over the angular coordinate $\phi_1$ of the wave vector $\bm k_1$
can be performed analytically.  The remaining seven integrals are computed
in one single Monte Carlo integration.
The wave-vector integrations need to be cut at some
$k_\mathrm{max}$.  The larger the choice of $k_\mathrm{max}$, the more
multipoles have to be considered when computing first-order results and
Green functions, increasing the demand for CPU time.  Fortunately it turns
out that the integrand decays quickly for $k_1$ or $k_2$ larger than
$k$. The $k_1$ dependence of the integrand is illustrated in
Fig.~\ref{fig:source},
depicting the suppression for large values of $k_1$. This
effect is due to the two primordial power spectra in the expectation
value  of the source terms,
Eq.~(\ref{Finaleq}). Since $k_2 = |\bm k - \bm k_1|$, for large $k_1$ the power
spectra suppress the integrand with $\mathcal O(k_1^{-6})$.  We find 
that choosing $k_\mathrm{max} = 0.3 \,\mathrm{Mpc}^{-1}$ is 
more than adequate, see Section~\ref{sec:stability}. 

\FIGURE[b]{
\epsfig{file=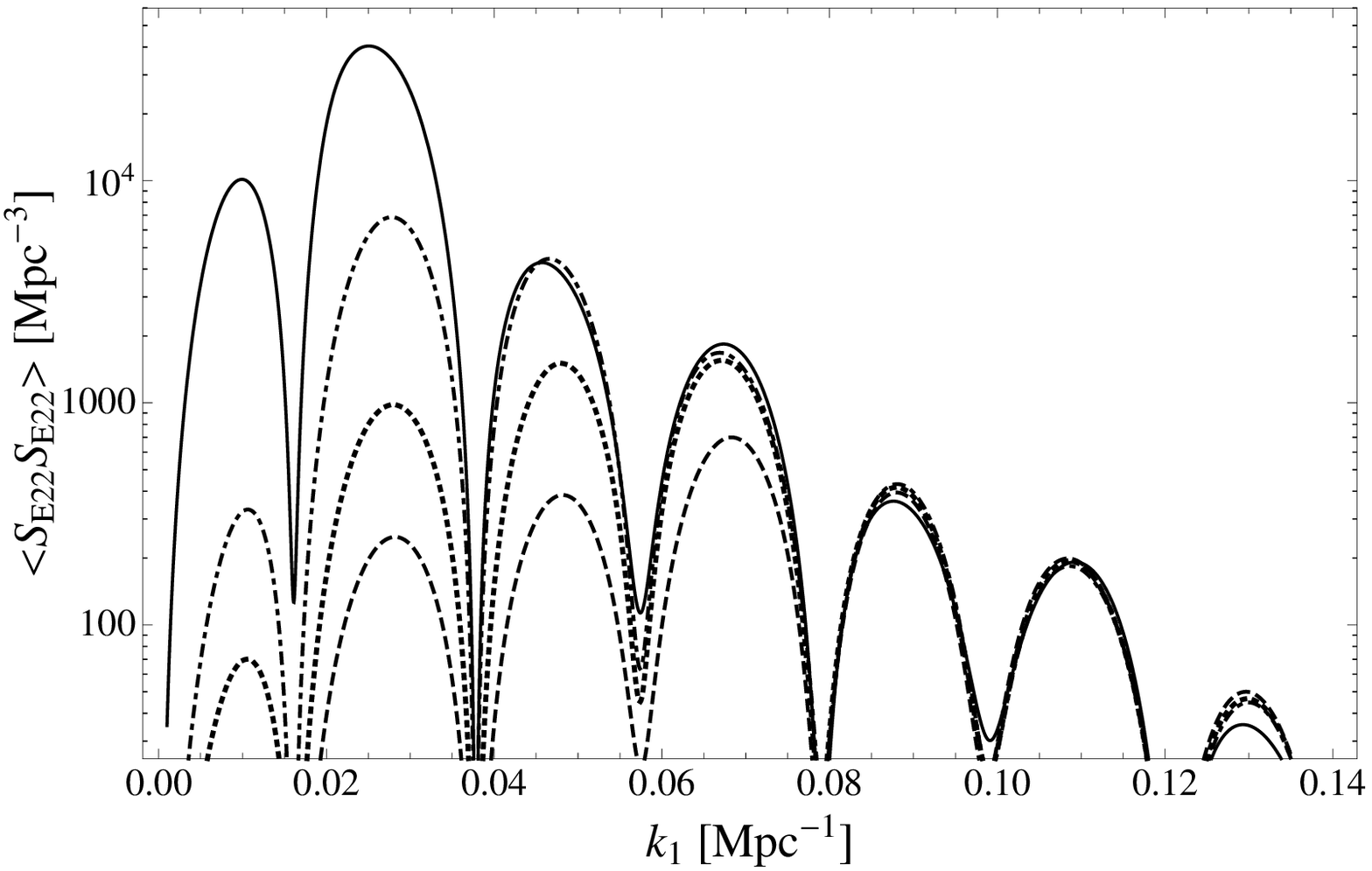,width=0.7\linewidth}
\caption{\label{fig:source}
The $k_1$ integrand of $\langle S_{E,22}(\bm k, \eta) S_{E,22}^*(\bm
k', \eta')\rangle$ (cf.
Eq.~\protect\eqref{Finaleq}) integrated over $\bm k'$.
In the convolution integral over
$\bm k_1$ we have only performed the integrals over the orientation, not
over the magnitude $k_1$ which is plotted on the abscissa,
for $\eta=\eta'=290 \,\mathrm{Mpc}/c$ and
(from top to bottom in the second peak) $k=0.02
\,\mathrm{Mpc}^{-1}$, $k=0.04 \,\mathrm{Mpc}^{-1}$, $k=0.06
\,\mathrm{Mpc}^{-1}$ and $k=0.08
\,\mathrm{Mpc}^{-1}$. The primordial power-spectrum is set to 
$P_\Phi(k) = 1/k^3$.  The main contribution is always located around
$k_1 \approx  k$. If $k_1$ is larger than $k$, the integrand is
suppressed by two primordial power spectra with large arguments.
This allows us to cut the $k_1$ integration at some $k_1$
sufficiently larger than $k$.}
}

Since we calculate only collisional sources which are multiplied by
the visibility function, many time integrals could be restricted to $200
\, \mathrm{Mpc}/c \lesssim \eta \lesssim 500 \,\mathrm{Mpc}/c$. This also 
applies to the Green function integrals, where the integrand is 
tight-coupling suppressed as discussed in Section~\ref{sec:green}. However, 
in the numerical results presented below, we use $\eta_\mathrm{ini}$ as lower 
bound for the time integrals. The adaptive Monte Carlo algorithm ({\sc Vegas})
\cite{Lepage:1977sw} automatically samples fewer points between
$\eta_\mathrm{ini}$ and $200 \,\mathrm{Mpc}/c$ than for later times.
The upper bound of the time integrations will be denoted by
$\eta_\mathrm{fs}$. To verify that
$\eta_\mathrm{fs} = 500 \,\mathrm{Mpc}/c$ is indeed a reasonable choice, we 
study the convergence of the result with increasing 
$\eta_\mathrm{fs}$, see Section~\ref{sec:stability}.

Only source terms $S_{X,lm}$ with small multipole moment $l$ have to be taken
into account, since higher first-order moments are only generated
by free streaming which takes time. For typical values of $k$, the octupole
reaches its first maximum only after $\eta_\mathrm{rec}$, so that its
contribution to the integrand which is peaked around recombination is small.
Higher moments are generated even later.
If Eq.~\eqref{eq:finalsolution} is used to compute the polarization modes,
there is an additional suppression of the intensity sources $S_{I,lm}$: since
the function $j_{nm}$ mixes only the two polarization modes, the intensity
source contributes to polarization solely via the Green functions in the
second line of Eq.~\eqref{eq:finalsolution}. But as stated in Section
\ref{sec:green}, a Green function $G_{mn}(\eta, \eta')$ for multipoles 
coupled only by the scattering term vanishes if the scattering rate
$|\dot\kappa|$ goes to zero. Accordingly, due to the additional
$|\dot\kappa|$ factor in front of the Green function, for small scattering 
rates the contribution of $S_{I,lm}$ sources is proportional to 
$|\dot\kappa|^2$ and
thus suppressed much earlier than other contributions proportional to
$|\dot\kappa|$. Hence, sources $S_{I,lm}$ with $l > 2$ are not only
tight-coupling suppressed at early times but in addition do not contribute
at late times.

In the following we denote by $n_l$ the highest source-term 
multipole moment included. All sources $S_{X,lm}$ with $l > n_l$ 
are neglected.  By the reasoning above we conclude that $n_l = 3$ 
should already provide reasonable results, since $S_{X,3m}$ is the 
highest source that contains the first-order quadrupole. 
This will be verified in detail in Section~\ref{sec:stability}.  Note
that the source $S_{X,n_lm}$ contains first-order solutions with $l = n_l + 1$,
i.e., for $n_l = 3$ we need the first order solutions up to $l = 4$.

We compute all first-order quantities and Green functions
in advance, before performing the integration. However, in the integrand
they are required as function of conformal time and wave number.
Thus we tabulate the solutions, storing their values for $n_k$ equidistant
wave-number values $k_0 = 0$, \dots, $k_{n_k - 1} = k_\mathrm{max}$
and $n_a$ time values $\eta_0 = \eta_\mathrm{ini}$, \dots,
$\eta_{n_a - 1} = \eta_\mathrm{fs}$. The time steps are chosen such
that the corresponding scale factors are equidistant. We obtain
one $n_k \times n_a$ array for each first order quantity and one
$n_k \times n_a \times n_a$ array for each Green function.
In the further computation, the tables are interpolated using basis splines
(B-splines). To obtain accurate results, the distance between successive 
$\eta$ and $k$ values has to be much smaller than the typical
scale of the interpolated function's fluctuations.


\subsection{Test in the tight-coupling regime}
\label{sec:testtight}
In order to test our numerics we compare the numerical results for the 
two-point function at early times before recombination with the analytically 
known second-order tight-coupling solution.  At early times polarization 
vanishes and the photon quadrupole is given by
$10 C^{-,2}_{m_1,m}v_{e,[m_1]}^{(1)}(\bm{k}_1)v_{e,[m_2]}^{(1)}(\bm{k}_2)$.
Table \ref{tab:tightc} shows the analytical and numerical values of the tensor
power-spectrum $P_{22,2}^{X (2)}(k)$ for $k= 0.05 \,\mathrm{Mpc}^{-1}$ at
conformal time $\eta = 200 \,\mathrm{Mpc}/c$.  For this test, the primordial
power-spectrum is set to $P_\Phi(k) = 1/k^3$. As expected, the 
polarization modes are strongly suppressed and the unpolarized
quadrupole has the correct size.  The small difference between the 
numerical values and the tight-coupling solution comes from the 
finite scattering rate limiting the validity of the tight-coupling 
solution for which an infinite scattering rate is assumed.  Evidently, 
in contrast to the unpolarized quadrupole, the E mode is strongly 
suppressed.  Even stronger suppressed is the B mode since it is not
coupled directly to the unsuppressed intensity quadrupole but only indirectly
via the E mode by the free-streaming term. 

\TABLE[t]{
\label{tab:tightc}
\begin{tabular}{|c|c|c|}
\hline
$\quad X\quad$ & tight-coupling solution & numerical value \\
\hline
\hline
$I$  & $8063 \pm 9$ & $8128 \pm 10$ \\
\hline
$E$ & $0$ & $14.6 \pm 2.9$ \\
\hline
$B$ & $0$ & $0.01 \pm 0.02$ \\
\hline
\end{tabular} 
\caption{Comparison of the tight-coupling solution and the numerical result
for the second-order power spectrum $P_{22,2}^{X(2)}(k)$;
$k = 0.05 \,\mathrm{Mpc}^{-1}$, $\eta = 200 \,\mathrm{Mpc}/c$,
primordial power-spectrum $P_\Phi(k) = 1 / k^3$.}
}

\subsection{Numerical stability tests}
\label{sec:stability}

The numerical error of our final result is made up of a systematic error
and the statistical error from the Monte Carlo integration. While for the
latter the {\sc Vegas} integration routine provides a reliable estimate,
quantifying the systematic error requires involved analysis.

\FIGURE[t]{
\epsfig{file = 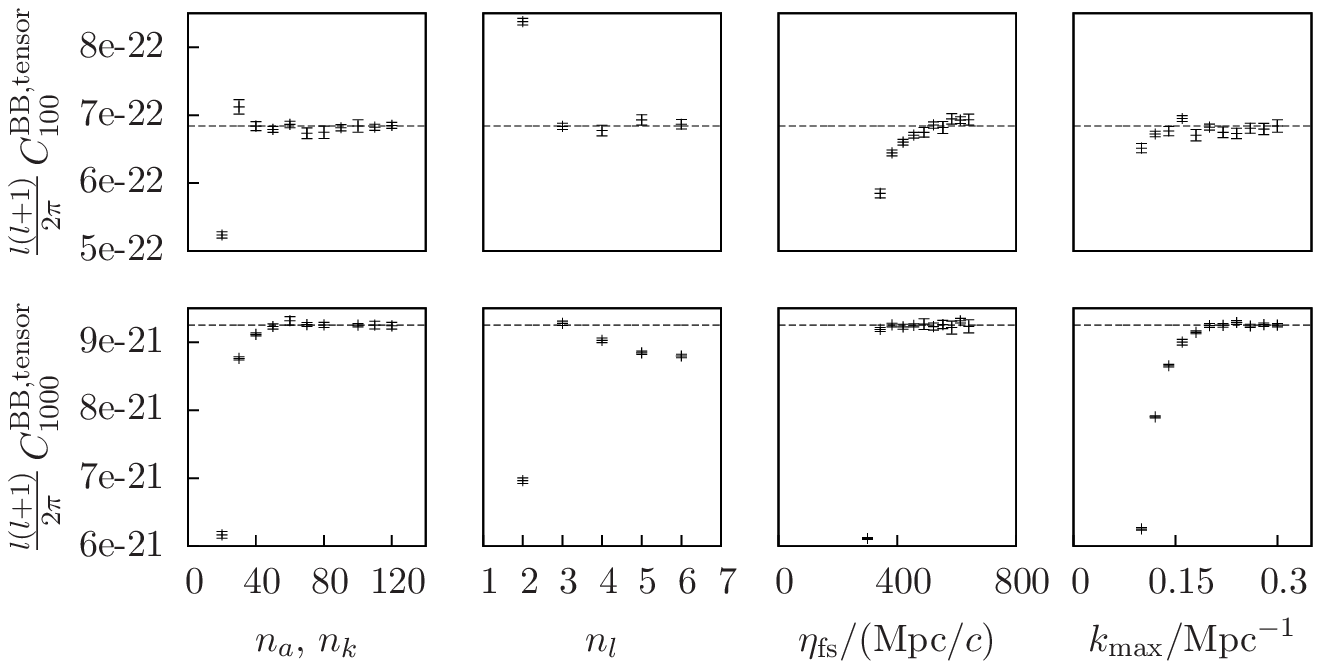, trim = 1cm 0 0 0}
\caption{\label{fig:conv}Dependence of the numerical result on the numerical
parameters. The tensor ($m = \pm 2$) contribution to the $C_l^{BB}$
value for $l = 100$ (top row) and for $l = 1000$ (bottom
row) is plotted vs. (from left to right) the number of scale-factor steps $n_a$
and wave-number steps $n_k$, the number of multipole moments taken into account
in the sources $S_n$, the scale factor $a_\mathrm{fs}$ from which we start
using the free-streaming solution, and the truncation of the wave-number
integrals $k_\mathrm{max}$. All parameters except for the parameter used as
abscissa are kept fixed at the values $n_a = n_k = 100$, $n_l = 3$,
$\eta_\mathrm{fs} = 500 \,\mathrm{Mpc}/c$ (corresponding to 
$a_\mathrm{fs} = 0.002064$)
and $k_\mathrm{max} = 0.3 \,\mathrm{Mpc}^{-1}$. The central value of the result
obtained using the default parameters is indicated by dashed lines.
}
}

The systematic error is controlled by five parameters: $\eta_\mathrm{fs}$,
$k_\mathrm{max}$, $n_a$, $n_k$ and $n_l$. 
Larger values for any of the five parameters yield better results but
require more CPU time. To balance both demands, we study the variation
of the numerical result under variation of the parameters. Values 
which yields a systematic error
on the few percent level are $n_a = n_k = 100$, $n_l = 3$,
$\eta_\mathrm{fs} = 500 \,\mathrm{Mpc}/c$
and $k_\mathrm{max} = 0.3 \,\mathrm{Mpc}^{-1}$.

\FIGURE[t]{
\hspace*{-1.5cm}
\includegraphics[width=11cm]{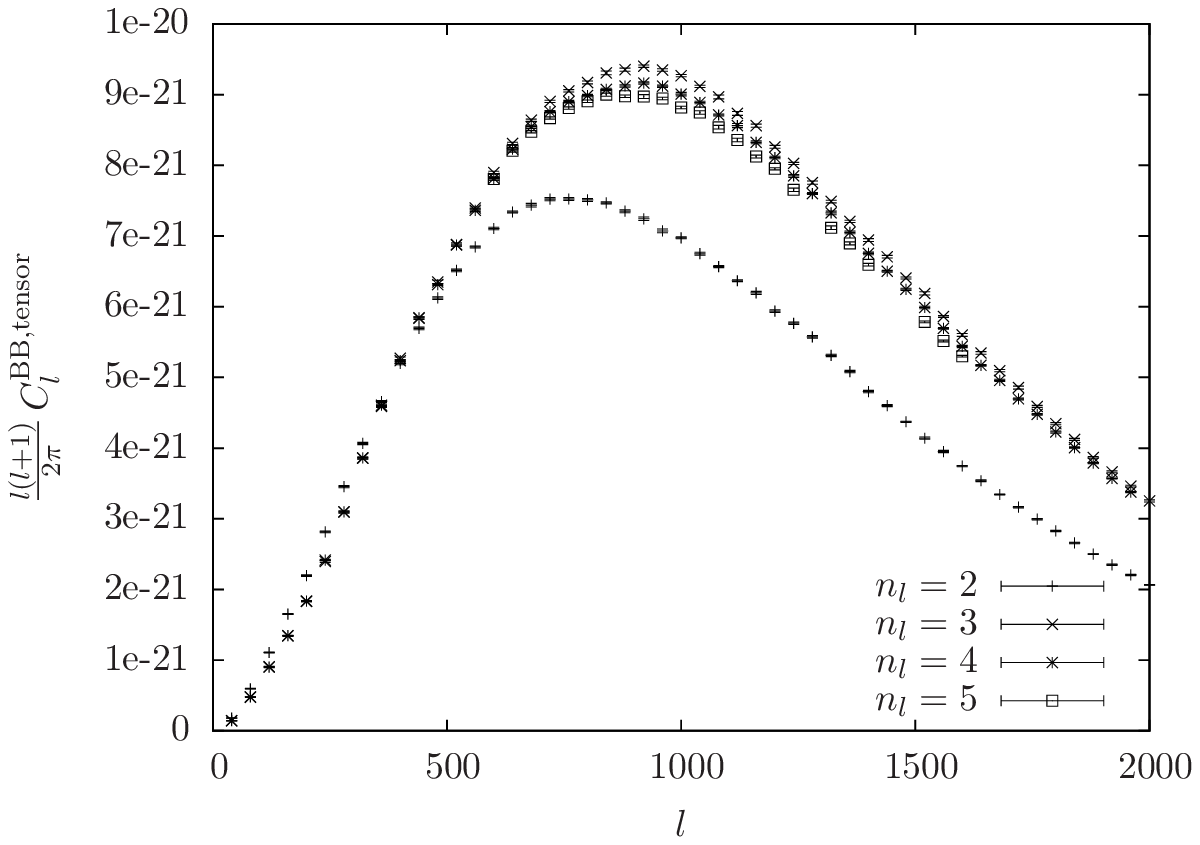}
\hspace*{-1.7cm}
\includegraphics[width=11.5cm]{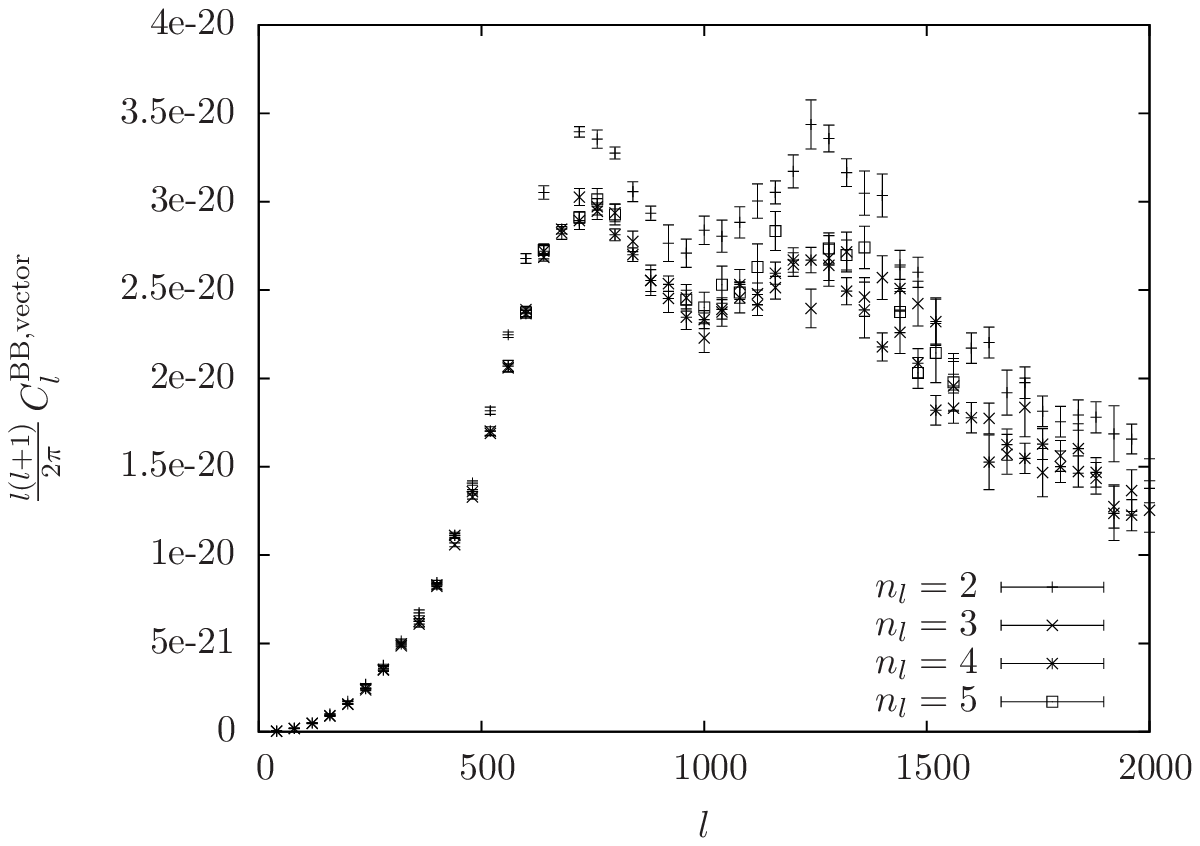}
\caption{\label{fig:clbb-conv} Tensor ($|m| = 2$, upper plot)  and 
vector contribution  ($|m| = 1$, lower plot) to the $C_l^{BB}$
spectrum, computed with different values of $n_l$, the number of
multipoles in the source terms.}
}

Figure~\ref{fig:conv} shows the stability of the result for the tensor
contribution ($m=\pm 2$) to the angular power spectrum computed using these
parameters.
In the figure we identify the choice of $\eta_\mathrm{fs}$ to dominate the
systematic error for small $l$, whereas for large $l$ neglecting
higher multipoles in the sources is the most relevant approximation.
This can
be easily understood by considering the $l$ dependence of the spherical
Bessel-function $j_l(k (\eta_0  - \eta))$ in the line-of-sight
integral.  For given $l$, the Bessel-function is small if $k (\eta_0  - \eta)
\lesssim l$ and oscillates for $k (\eta_0  - \eta) \gtrsim l$; the main
contribution to the integral comes from $k (\eta_0  - \eta) \approx l$.  Since
$\eta_0 \gg \eta_\mathrm{fs} \geq \eta$, the integral is therefore dominated by
contributions at $k \approx l / \eta_0$.  Accordingly, the 
precondition for free
streaming, $|\dot\kappa| \ll k$, is fulfilled at later times for small $l$
than for large $l$. Consequently, the computation for small $l$ requires
a large $\eta_\mathrm{fs}$. With increasing $l$, the main contribution 
to the integral moves to larger $k$ values.  While for small $k$ the 
lowest multipole moments of the first-order result are largest, 
with increasing $k$ they are superseded by the higher moments.
This can already be noticed in Fig.~\ref{fig:conv}, where increasing the 
number of multipole moments in the source terms from $n_l = 3$ to $n_l = 4$
or $n_l = 5$ changes the result significantly for $l = 1000$, but not 
for $l = 100$. However, Fig.~\ref{fig:clbb-conv} shows that for both,
the tensor and the vector contribution to the $C_l^{BB}$ B-mode 
angular power spectrum this effect
does not gain importance for even larger $l$: the spectra for $n_l = 3$ are
close to the ones computed with $n_l = 4$ and 5 for all $l$, whereas $n_l = 2$
is clearly insufficient to cover the entire range up to 
$l\approx 2000$.

\FIGURE[t]{
\hspace*{-1.5cm}
\includegraphics[width=11cm]{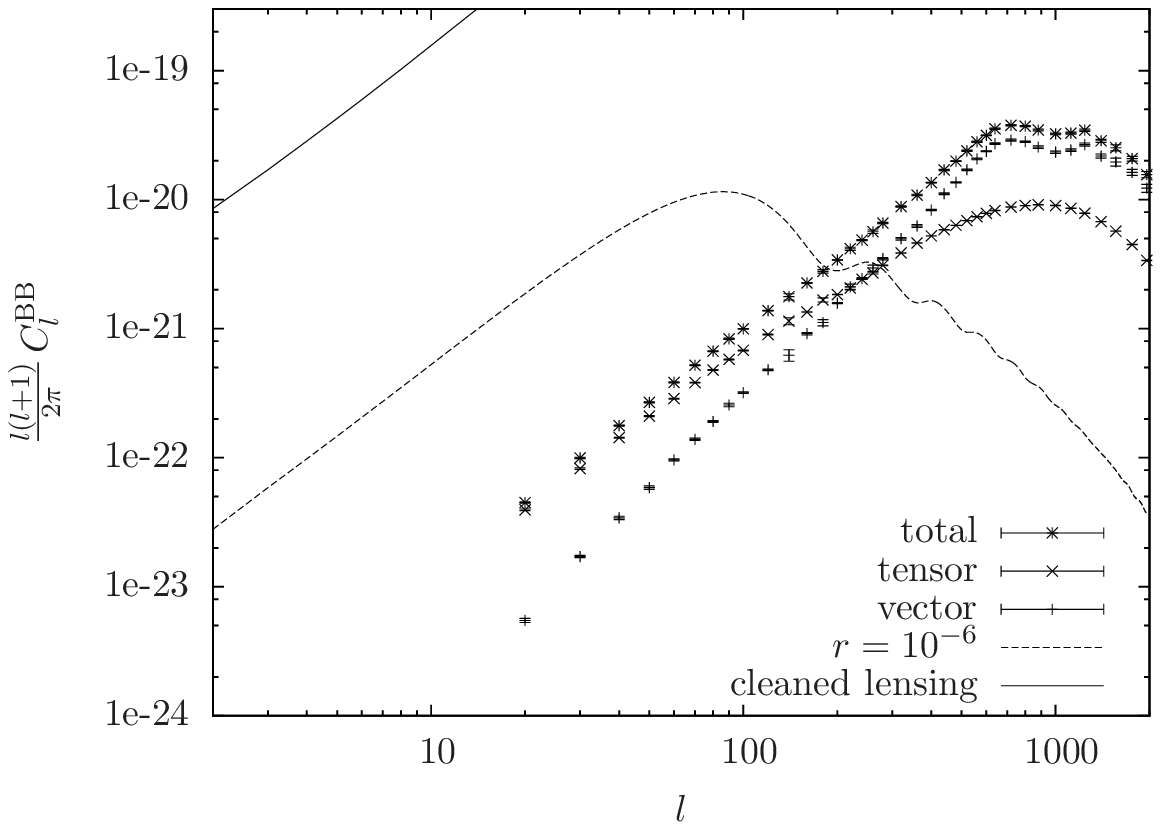}
\hspace*{-1.6cm}
\includegraphics[width=11.3cm]{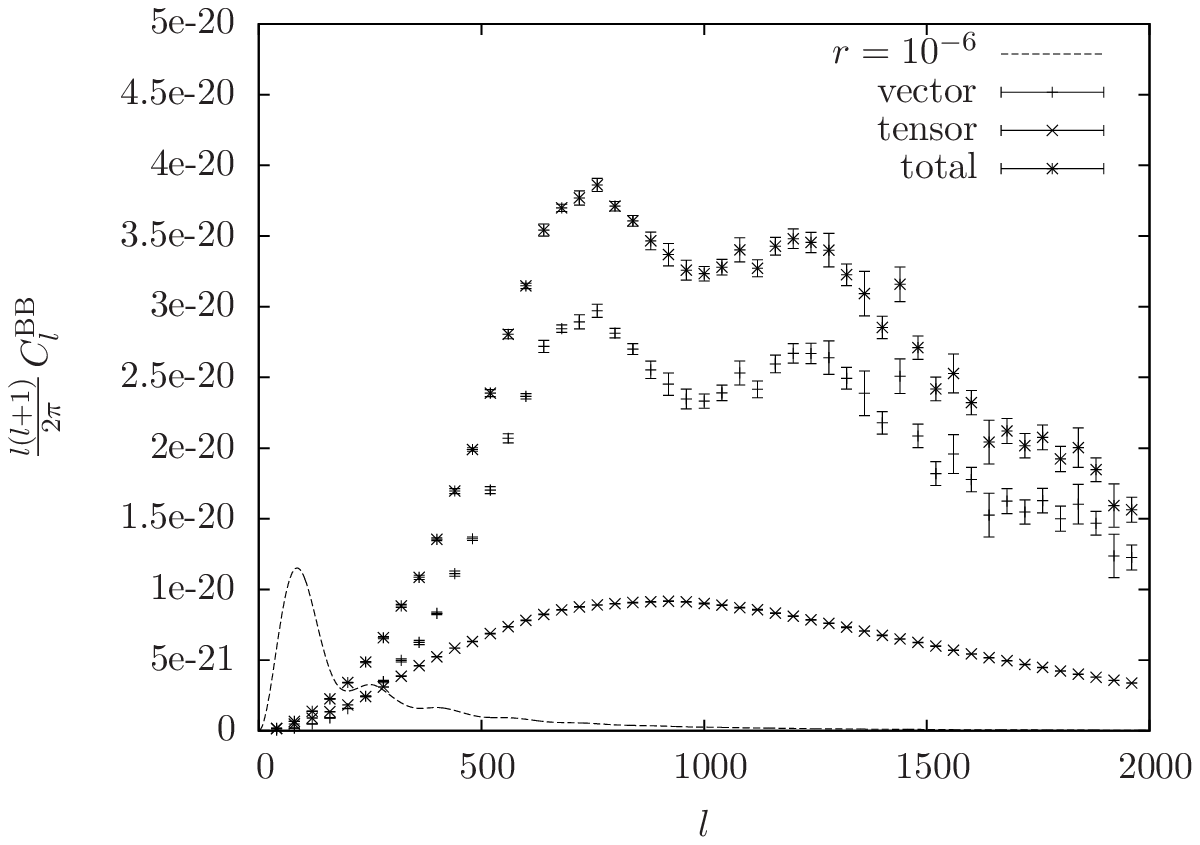}
\caption{
\label{fig:final}
The $C_l^{BB}$ angular power spectrum from second-order 
scattering sources (points) compared to the spectrum induced by  
primordial tensor perturbations with tensor-to-scalar ratio 
$r = 10^{-6}$ (dashed line) and the weak-lensing signal 
cleaned by a factor 40 (solid line, upper panel only).  
We display separately the second-order contributions from 
vector ($|m|=1$) and tensor ($|m|=2$) perturbations, and 
their sum. The upper and lower plot show the same numerical 
data, first on a logarithmic, then on a linear scale.  
(Numerical parameters are fixed to the values $n_a = n_k = 100$, $n_l = 4$,
$\eta_\mathrm{fs} = 500 \,\mathrm{Mpc}/c$,
and $k_\mathrm{max} = 0.3 \,\mathrm{Mpc}^{-1}$.)}
}


\section{Result and conclusion}
\label{sec:results}

We summarize our results by showing the B-mode angular power 
spectrum $C_l^{BB}$ from the second-order collision term in 
Fig.~\ref{fig:final} both, on a logarithmic (upper plot) and on a 
linear scale (lower plot). Similar to the weak-lensing 
induced contribution, the angular power spectrum grows 
rapidly with $l$ until it reaches a maximum around 
$l\approx 1000$. At $l \lesssim 200$ the tensor mode 
contribution ($|m|=2$) is larger than the vector ($|m|=1$) 
contribution, but the latter dominates in the peak region. 
The vector mode also shows a double peak structure 
around $l\approx 1000$. 

We find that for $|m| = 1$ the source terms $S_n$ are dominated by the
terms containing a product of the first-order electron velocity 
$v_{e,[m]}^{(1)}$ and radiation dipole $\Delta_{I,1m}^{(1)}$ 
(closely related to 
the electron velocity in tight coupling) with each other 
or themselves.  Setting all other first-order quantities in $S_n$ to
zero, only a few terms remain. The source term $S_{I,lm}$ 
consists of only the fourth and seventh line of the right-hand side 
of Eq.~\eqref{eq:2ndorderI}, the source $S_{E,lm}$ 
of the third and fourth line of the right-hand side 
of Eq.~\eqref{eq:2ndorderE}, and the sources
$S_{B,lm}$ and $S_{v_e,1m}$ vanish altogether as can
be seen from Eq.~\eqref{eq:2ndorderB} and
Eq.~\eqref{eq:2ndorderv}. With these approximations, we obtain 
the double-peaked spectrum with the first and second peak
reaching more than 80\% and 60\%, respectively, of the
corresponding peaks in the full result,
Fig.~\ref{fig:final}.  This indicates that the double-peak
structure is related to the evolution of the first-order electron
velocity/radiation dipole.  The electron velocity oscillates over time until
recombination and thus throughout the time interval relevant
for the integration.  The frequency depends on the wave
number which in the source terms is $k_1$ and $k_2$,
respectively. For the vector-mode contribution the wave-vector 
convolution integral is dominated by contributions at
$k_1 \approx k$ and $k_2$ small or vice versa and the $k$
integral by contributions at $k \approx l / \eta_0$. Since
$\eta_0 \approx 14000\,\mathrm{Mpc}/c$, this relates the $l$
values of the two peaks, $l \approx 750$ and $l \approx
1250$, to the wave numbers $k \approx
0.05\,\mathrm{Mpc}^{-1}$ and $k \approx
0.09\,\mathrm{Mpc}^{-1}$.  For both $k$ values, the
oscillation of the electron velocity reaches an extremum at
recombination, where for $k \approx 0.09\,\mathrm{Mpc}^{-1}$
it has undergone one full oscillation more than for $k
\approx 0.05\,\mathrm{Mpc}^{-1}$.  The intermediate wave
number $k \approx 0.07\,\mathrm{Mpc}^{-1}$, for which the
oscillation stops at recombination with a phase shift of
only half a period, corresponds to $l \approx 1000$.  At
this $l$ value we observe a local minimum of the spectrum.
The tensor spectrum does not show a double-peak structure, since a
larger range of wave numbers contributes to the convolution
integral so that any oscillations are averaged out after
integration.

The amplitude of collision-induced B-mode polarization is
several orders of magnitude smaller 
than the weak-lensing signal. At $l\approx 200$ it 
is comparable to the B-mode power generated by primordial 
gravitational waves with a tensor-to-scalar ratio 
$r\approx 10^{-6}$ (shown as dashed line in the Figure), 
growing to $2\cdot 10^{-4}$ at $l\approx 1000$. 
The small amplitude of the collisional contribution is 
presumably due to the fact that the sources involve 
only the cross-coupling of the first-order photon perturbations 
to the velocity of the baryon fluid and further 
are mainly localized in time to the recombination era. It may 
therefore be of interest to include the reionization 
history into the computation to obtain a realistic estimate 
of the power spectrum at low values $l\approx 10$.
 
The collision-induced spectrum may be compared to the 
spectrum induced by second-order vector and 
tensor metric perturbations \cite{Mollerach:2003nq}.  
These act as sources for vector and tensor perturbations 
of the radiation intensity, which can then be 
converted into E-mode polarization by Thomson scattering 
through the first-order collision term, and 
subsequently to B-mode polarization by free-streaming.
We find that both contributions, the present one and 
the one from metric perturbations, are similar in 
magnitude. If the weak-lensing signal could be ``cleaned'' 
completely, these second-order terms would constitute 
a background to the search for primordial gravitational 
waves at the level of $r < 10^{-6} \ldots 2\cdot 10^{-4}$ (depending 
on $l$), which represents a challenge to 
CBR polarization experiments.

\section*{Acknowledgement}

We thank E.~Dimastrogiovanni and C.~Pitrou for helpful discussions 
and communications.  
This work is supported in part by the Gottfried Wilhelm 
Leibniz programme of the Deutsche Forschungsgemeinschaft (DFG)  and the 
DFG Graduiertenkolleg ``Elementar\-teil\-chen\-physik an der TeV-Skala''.

\appendix

\section{Summary of equations}
\label{ap:equations}

In this appendix we summarize further equations related to the content 
of the main text. For conventions and notation not stated explicitly 
we refer to Ref.~\cite{Beneke:2010eg}. The equations are given in
Fourier space 
conjugate to position, vectors and tensors are decomposed into spherical 
components. Phase space densities are expanded into multipoles with 
respect to the photon momentum direction.

In the absence of first-order vector and 
tensor perturbations the first-order Boltzmann hierarchy 
for the energy-integrated photon phase space density multipoles 
(\ref{eq:Deltadef}) is given in conformal Newtonian gauge by 
\begin{eqnarray}
&& \frac{\partial}{\partial \eta}\Delta_{I,lm}^{(1)}(\bm{k}) + 
\sum_{\pm} (\mp i) \Delta_{I,(l \pm 1)m_1}^{(1)}(\bm{k})
k^{[m_2]}C_{m_1 m}^{\pm,l} +
4 \delta_{l0}\dot{D}^{(1)}(\bm{k}) + 
4 \delta_{l1}i k^{[m]}A^{(1)}(\bm{k}) 
\nonumber \\
&&\hspace*{0.5cm} = 
|\dot{\kappa}|\left\{-\Delta_{I,lm}^{(1)}(\bm{k}) + 
\delta_{l0}\Delta_{I,00}^{(1)}(\bm{k})  + 
4\delta_{l1} v^{(1)}_{e,[m]}(\bm{k}) + 
\delta_{l2}\frac{1}{10}\left(\Delta_{I,2m}^{(1)}(\bm{k}) - 
\sqrt{6}\Delta_{E,2m}^{(1)}(\bm{k}) \right)\right\},
\nonumber \\[0.2cm]
&& \frac{\partial}{\partial \eta}\Delta_{E,lm}^{(1)}(\bm{k}) + 
\sum_{\pm} (\mp i) \Delta_{E,(l \pm 1)m_1}^{(1)}(\bm{k})
k^{[m_2]}D_{m_1 m}^{\pm,l} - i \Delta_{B,l m_1}^{(1)}(\bm{k}) 
k^{[m_2]} D^{0,l}_{m_1 m}  
\nonumber \\
&&\hspace*{0.5cm} =
|\dot{\kappa}|\,\bigg\{-\Delta_{E,lm}^{(1)}(\bm{k}) -
\delta_{l2} \frac{\sqrt{6}}{10} \left(\Delta_{I,2m}^{(1)}(\bm{k}) - 
\sqrt{6}\Delta_{E,2m}^{(1)}(\bm{k}) \right)\bigg\}, 
\nonumber \\[0.2cm]
&& \frac{\partial}{\partial \eta}\Delta_{B,lm}^{(1)}(\bm{k}) + 
\sum_{\pm} (\mp i) \Delta_{B,(l \pm 1)m_1}^{(1)}(\bm{k}) 
k^{[m_2]} D^{\pm,l}_{m_1 m}  + 
i \Delta_{E,l m_1}^{(1)}(\bm{k}) k^{[m_2]}D_{m_1 m}^{0,l} 
\nonumber \\
&&\hspace*{0.5cm} =
|\dot{\kappa}|\,\bigg\{ - \Delta_{B,lm}^{(1)}(\bm{k})\bigg\}.
\end{eqnarray}
Since B-mode polarization is not generated at first order from scalar 
perturbations, we may set $\Delta_{B,lm}^{(1)}(\bm{k})=0$ in these 
equations. The Boltzmann equations for the massless neutrino variables 
$\Delta_{\nu,lm}^{(1)}(\bm{k})$ are the same 
as the equation for $\Delta_{I,lm}^{(1)}(\bm{k})$ with the collision 
term on the right-hand side set to zero. 
Baryons and electrons form a tightly coupled fluid with 
first-order variables $\delta_b^{(1)}(\bm{k})=
\left[\delta \rho_b/\bar \rho_b\right]^{(1)}\!(\bm{k})$, 
$v_{e,[m]}^{(1)}(\bm{k})$ determined by 
\begin{eqnarray}
&& \frac{\partial}{\partial \eta}\delta_b^{(1)}(\bm{k}) -
\sum_{m=-1}^1 (-1)^{m} 
 i k^{[m]} v^{(1)}_{e,[-m]}(\bm{k}) + 3\dot D^{(1)}(\bm{k}) = 0, 
\nonumber\\[0.2cm]
&& \left(\frac{\partial}{\partial \eta} +H_C\right) 
v^{(1)}_{e,[m]}(\bm{k}) +i k^{[m]}A^{(1)}(\bm{k})
=-\frac{|\dot{\kappa}|}{4 R}
\left\{4 v^{(1)}_{e,[m]}(\bm{k}) - \Delta_{I,1m}^{(1)}(\bm{k})\right\}
\quad
\end{eqnarray}
with $R=3 \bar\rho_b/(4 \bar\rho_\gamma)$.
The cold dark matter perturbations 
$\delta_c^{(1)}(\bm{k})=\left[\delta \rho_c/\bar 
\rho_c\right]^{(1)}\!(\bm{k})$, 
$v_{c,[m]}^{(1)}(\bm{k})$ are described by the same equations 
with the collision term on the right-hand side set to zero. 
The system closes with the Einstein equations for the first-order 
metric perturbations, which can be put into the form
\begin{eqnarray}
&& k^2 D^{(1)}(\bm{k}) + 3H_C \dot{D}^{(1)}(\bm{k}) -3 H_C^2 A^{(1)}(\bm{k}) 
= 4\pi G a^2 \,\delta\rho^{(1)}(\bm{k}),
\nonumber \\[0.2cm]
&& C_{m_1 m}^{-,2} k_{[m_1]} k_{[m_2]}\, 
(A^{(1)}+D^{(1)})(\bm{k}) = 8\pi G a^2 \alpha_m\Sigma_{[m]}^{(1)}(\bm{k}).
\end{eqnarray}
The density perturbation and anisotropic stress 
are given by
\begin{eqnarray}
\delta \rho^{(1)}(\bm{k}) &=& \bar \rho_b\delta_b^{(1)}+ 
\bar \rho_c\delta_c^{(1)} + 
\bar \rho_\gamma\Delta_{I,00}^{(1)}+\bar \rho_\nu\Delta_{\nu,00}^{(1)},
\nonumber\\[0.2cm]
\Sigma_{[m]}^{(1)}(\bm{k}) &=& \frac{2}{15\alpha_m}\left(\bar \rho_\gamma 
\Delta_{I,2m}^{(1)}+\bar \rho_\nu\Delta_{\nu,2m}^{(1)}\right).
\end{eqnarray}
In practice, we solve the first-order equations only for mode vectors 
$\bm{k}$ aligned with the 3-axis, in which case $k^{[0]}=ik$ and 
$k^{[\pm 1]}=0$. In the absence of vector and tensor modes the 
$m=\pm1,\pm2$ components all vanish for this choice of $\bm{k}$. 
The solution for general $\bm{k}$ needed in the source term of the 
second-order equations is obtained from the $m=0$ solution for 
$\bm{k}=k\bm{e}_3$ by a rotation. In general, if $T_{lm}(\bm{k})$ 
is a spherical tensor of rank $l$, the relation between its 
components for $\bm{k} = k \bm{\hat k}$ and $\bm{k} = k \bm{e}_3$ 
is 
\begin{equation}
T_{lm}(\bm{k}) = 
\sum_{m'} \,T_{lm'}(k \bm{e}_3) \,
D^{(l)}_{m' m}(R^{-1}) = 
\sqrt{\frac{4\pi}{2 l+1}}\,\sum_{m'} T_{lm^\prime}(k \bm{e}_3)\,
Y^{-m^\prime}_{lm}(\bm{\hat k})\,
\label{eq:Tlmrotation}
\end{equation}
where $D^{(l)}_{m' m}(R^{-1})_{m' m}$ denotes the Wigner 
function for the rotation $R^{-1}$ with $\bm{\hat k}=R\bm{e}_3$, 
which can be expressed in terms of the spin-weighted 
spherical harmonics \cite{Goldberg1967}. 
We apply this to $A^{(1)}$, $D^{(1)}$, $\Delta^{(1)}_{I,00}$, 
$\delta_b^{(1)}$ etc. with $l=0$, to  
$\Delta^{(1)}_{I,1m}$, $v^{(1)}_{e,[m]}$ etc. with $l=1$ 
and to $\Delta^{(1)}_{X,lm}$ and  $\Delta^{(1)}_{\nu,lm}$ with 
any $l$. In this case, due to the absence of first-order vector and 
tensor perturbations, only the $m'=0$ term contributes to the sum in
Eq.~(\ref{eq:Tlmrotation}). 

Since we do not consider polarization induced by the second-order 
vector and tensor metric perturbations, themselves induced by 
the first-order scalar modes, we do not need to solve the second-order 
Einstein equations.  
The only second-order quantity other than the 
photon perturbations that we must solve for is the electron velocity
for which we use the truncated equation (\ref{eq:2ndorderv}) as explained 
in the main text. For completeness we provide here the complete 
equations for the second-order variables of the baryon-electron 
fluid. We define the fluid variables through the energy-momentum 
tensor in the local inertial frame at rest and aligned with 
the general coordinate system
\begin{eqnarray}
T_{AB} &=& [e_A]^\mu [e_B]^\nu \,T_{\mu\nu} 
= (\rho+p) u_A u_B - p \,\eta_{AB} +\Sigma_{AB}
\nonumber\\[0.2cm]
&=&\int \frac{d^3{\bm p}}{(2\pi)^3}\,g(\eta,\bm{x},\bm{p})\,
\frac{p_A p_B}{E}.
\end{eqnarray}
Here $p^A$ denotes the physical (not co-moving) particle 
momentum in the local frame and $g(\eta,\bm{x},\bm{p})$ the 
phase-space distribution, for which we take a local Maxwell-Boltzmann 
distribution. The four-velocity is parameterized in the form 
$u^A=(1/\sqrt{1-\bm{v}^2},\bm{v})$. We then calculate the 
conformal-time derivative of the expression in the second line 
using the results of Ref.~\cite{Beneke:2010eg}. Together with 
\begin{equation}
T_{00} = \rho + \bar\rho \,[\bm{v}^{(1)}]^2 + \ldots, 
\qquad 
T_{0i} = - \rho v^i+\ldots,
\end{equation}
valid to second order in perturbations, we obtain the desired 
equations of the density and velocity perturbation of the 
baryon-electron fluid to second order:
\begin{eqnarray} \nonumber
&& \frac{\partial}{\partial \eta}\delta_b^{(2)}(\bm{k}) 
 - \sum_{m=-1}^1 (-1)^{m} 
 i k^{[m]} v^{(2)}_{e,[-m]}(\bm{k}) + 3\dot D^{(2)}(\bm{k})
\nonumber \\
&& \hspace*{1cm} 
-\sum_{m=-1}^1 (-1)^{m}\,
\bigg(ik^{[m]} \delta_b^{(1)} +
ik_2^{[m]}(A^{(1)}-D^{(1)})+2ik_1^{[m]}D^{(1)}\bigg)(\bm{k}_1)
\,v_{e,[-m]}^{(1)}(\bm{k}_2)  
\nonumber\\[0.2cm] 
&& \hspace*{1cm} +\, \sum_{m=-1}^1 (-1)^{m} H_C\,
v_{e,[m]}^{(1)}(\bm{k}_1) v_{e,[-m]}^{(1)}(\bm{k}_2)
- 6 D^{(1)}(\bm{k}_1)\dot{D}^{(1)}(\bm{k}_2)
+3\dot{D}^{(1)}(\bm{k}_1) \delta_b^{(1)}(\bm{k}_2)
\nonumber\\[0.2cm]
&& \hspace*{0.5cm} 
=\,-\frac{|\dot{\kappa}|}{4 R}
\sum_{m=-1}^1 (-1)^{m}v^{(1)}_{e,[-m]}(\bm{k}_1)
\left(4v^{(1)}_{e,[m]}-\Delta_{I,1m}^{(1)}\right)(\bm{k}_2),
\label{eq:deltab2}
\\[0.3cm]
&& \left(\frac{\partial}{\partial \eta}+H_C\right)
(v^{(2)}_{e,[m]}-B^{(2)}_{[m]})(\bm{k}) +i k^{[m]}A^{(2)}(\bm k) 
\nonumber \\[0.2cm] 
&& \hspace*{1cm} 
- \,ik_1^{[m]}A^{(1)}(\bm{k}_1)(A^{(1)}+D^{(1)})(\bm{k}_2) + 
\dot{D}^{(1)}(\bm{k}_1)v^{(1)}_{e,[m]}(\bm{k}_2) 
\nonumber \\
&& \hspace*{1cm} - \sum_{m'=-1}^1 
(-1)^{m'} i k_2^{[-m']} \,v^{(1)}_{e,[m']}(\bm{k}_1)v^{(1)}_{e,[m]}(\bm{k}_2) 
 \nonumber\\[0.2cm]
&&\hspace*{0.5cm}
=\,-\frac{|\dot{\kappa}|}{4 R}\,
\bigg\{4 v_{e,[m]}^{(2)}(\bm{k})- \Delta_{I,1m}^{(2)}(\bm{k})
+\,\bigg(A^{(1)} + \left[\frac{\delta x_e}{x_e}\right]^{(1)}\,\bigg)(\bm{k}_1)
\left(4 v^{(1)}_{e,[m]}(\bm{k}_2)-\Delta_{I,1m}^{(1)}(\bm{k}_2)\right)
\nonumber\\[0.2cm] 
&& \hspace*{1cm}
- \,v^{(1)}_{e,[m_2]}(\bm{k}_1)\Delta_{I,2m_1}^{(1)}(\bm{k}_2) 
C^{+,1}_{m_1 m}+ 4 v^{(1)}_{e,[m]}(\bm{k}_1)
\Delta_{I,00}^{(1)}(\bm{k}_2) \bigg\}.
\label{eq:ve2}
\end{eqnarray}
The left-hand sides of these equations are obtained from the 
$l=0$ and $l=1$ moments of the Boltzmann equation for massive particles 
given in \cite{Beneke:2010eg}. The right-hand sides follow from the 
collision term for photons using the tight coupling of baryons and electrons 
through Coulomb scattering and energy-momentum conservation in 
photon-electron Compton scattering. Eqs.~(\ref{eq:deltab2}) and 
(\ref{eq:ve2}) agree with the corresponding results in 
Ref.~\cite{Pitrou:2010sn}.

Finally, the coupling 
coefficients $C_{m_1m_2}^{\pm,l}$ and $D_{m_1m_2}^{0,l}$ that appear 
in the Boltzmann equations read
\begin{eqnarray}
C^{+,l}_{m\pm 1,m} &=& 
-\frac{\sqrt{(l+1\pm m)(l+2\pm m)}}{\sqrt{2}(2l+3)} 
\nonumber \\
C^{+,l}_{m,m} &=& \frac{\sqrt{(l+1)^2-m^2}}{2l+3} 
\nonumber \\
C^{-,l}_{m\pm 1,m} &=& \frac{\sqrt{(l-1\mp m)(l\mp m)}}{\sqrt{2}(2l-1)} 
\nonumber \\
C^{-,l}_{m,m} &=& \frac{\sqrt{l^2-m^2}}{2l-1} 
\nonumber \\
D^{0,l}_{m\pm 1,m} &=& \mp \frac{\sqrt{2(l+1\pm m)(l \mp m)}}{l (l+1)} 
\nonumber \\
 D^{0,l}_{m,m} &=& -\frac{2m}{l (l+1)}.
\label{eq:couplingcoeffs}
\end{eqnarray}

\section{Derivation of the line-of-sight solution}
\label{sec:Los}

The line-of-sight solution of the first-order Boltzmann equations have been
derived in Ref.~\cite{Seljak:1996is} (see also Ref.~\cite{Hu:1997hp}). 
We provide here a generalization which allows for more general
source terms including higher multipole moments and polarization modes.

Our aim is to solve the equation
\begin{equation}
  \dot{\Delta}_{n} + k C_{nm}\Delta_{m}  = -|\dot{\kappa}|\Delta_{n} +\rho_n
\end{equation}
for $\bm{k}=k \bm{e}_3$.
First we solve the homogeneous differential equation without the source term
$\rho_n$.  This can be done easily after reversing the multipole decomposition.
Then the homogeneous equation takes the simple form:
\begin{equation}
\dot{\Delta}_{ab} + i {\bm n}\cdot \bm k\,\Delta_{ab}
= -|\dot{\kappa}|\Delta_{ab},
\end{equation}
where $ab=++,+-,-+,--$ are the polarization indices. 
This equation is solved by
\begin{equation}
\Delta_{ab}(\eta) =
e^{-i {\bm n} \cdot \bm{k}\,(\eta-\eta')-\kappa(\eta, \eta')}\,
\Delta_{ab}(\eta'),
\end{equation}
where $\kappa(\eta, \eta')$ is the integral over $|\dot\kappa|$ 
from $\eta'$ to $\eta$. If the first argument of $\kappa$ is the
conformal time today $\eta_0$, we omit this argument in the following
and in the main text, and write $\kappa(\eta)= \kappa(\eta_0, \eta)$.
Next, we decompose this solution into multipoles. 
For this purpose, we expand the exponential in spherical harmonics and 
$\Delta_{ab}(\eta')$ in spin-weighted spherical harmonics, 
\begin{align}
\Delta_{ab}(\eta)
=&
\sum_{l_1} e^{-\kappa(\eta, \eta')}\,
(-i)^{l_1}\sqrt{4\pi(2l_1+1)}j_{l_1}(k(\eta-\eta'))
Y_{l_10}({\bm n})
\nonumber \\&\times
\sum_{l_2,m} i^{-l_2} \sqrt{\frac{4 \pi}{2 l_2 + 1}}\Delta_{ab,l_2m}(\eta')
Y_{l_2m}^s(\bm n),
\end{align}
and apply on both sides $i^l\sqrt{\frac{2l+1}{4\pi}}\int d\Omega\,
Y_{lm}^{s*}({\bm n})$ to obtain
\begin{eqnarray}
\nonumber
\Delta_{ab,lm}(\eta)
&=&
\sum_{l_1, l_2}
e^{-\kappa(\eta, \eta')}
\,i^{l-l_1-l_2}\frac{(2l+1)(2l_1+1)}{2l_2+1}
j_{l_1}(k(\eta-\eta'))\\
&&
\times\,\left(\begin{array}{ccc}l & l_1 &l_2 \\ m & 0 & m\end{array}\right)
\left(\begin{array}{ccc}l&l_1&l_2\\-s&0&-s\end{array}\right)
\Delta_{ab,l_2 m}(\eta').
\end{eqnarray}
The spin $s$ is zero for $ab = ++, --$ and $\pm 2$ for
$ab = \pm\mp$. Finally, instead of the photon helicity state densities,
we use the Stokes parameter distributions
\begin{eqnarray}
\Delta_{I,lm} &=& \frac{1}{2} (\Delta_{++,lm} + \Delta_{--,lm}),
\\
\Delta_{E,lm} &=& \frac 1 2 (\Delta_{+-,lm} + \Delta_{-+,lm}),
\label{eq:deltae} \\
\Delta_{B,lm} &=& \frac i 2 (\Delta_{+-,lm} - \Delta_{-+,lm}).\label{eq:deltab}
\end{eqnarray}
On the right-hand side of Eq.~\eqref{eq:deltae} and Eq.~\eqref{eq:deltab}
there is a sum and a difference, respectively, of Clebsch-Gordan coefficients
with opposite spin.  They are related by
\begin{equation}
\left(\begin{array}{ccc}l&l_1&l_2\\-s&0&-s\end{array}\right)
= (-1)^{l-l_1-l_2}\left(\begin{array}{ccc}l&l_1&l_2\\s&0&s\end{array}\right).
\end{equation}
Depending on the parity of $l-l_1-l_2$ we obtain either a sum or a difference
of the $+-$ and $-+$ components. For odd parity, this leads to a mixing of E
and B modes.
The resulting mode coupling can be written in a compact form
as matrix $H_{XX'}$ defined by $H_{XX'}(\text{even})=
\delta_{XX'}$ and $H_{XX'}(\text{odd})=\delta_{XI}\delta_{X'I} + i
\delta_{XE}\delta_{X'B} -i\delta_{XB}\delta_{X'E}$. Then 
\begin{eqnarray}\nonumber
\Delta_{X,lm}(\eta) &=&
\sum_{l_1, l_2} \sum_{X' = I,E,B}
e^{-\kappa(\eta, \eta')}
\,i^{l-l_1-l_2} \frac{(2l+1)(2l_1+1)}{2l_2+1}j_{l_1}(k(\eta-\eta'))\\
&&
\times\,\left(\begin{array}{ccc}l & l_1 &l_2 \\ m & 0 & m\end{array}\right)
\left(\begin{array}{ccc}l&l_1&l_2\\F_X&0&F_{X'}\end{array}\right)
H^*_{XX'}(l-l_1-l_2)\Delta_{X',l_2 m}(\eta'),
\label{eq:los1}
\end{eqnarray}
where $F_I = 0$, $F_E = F_B = -2$.
A more detailed explanation of the effect which introduces the matrix $H$ and
the properties of spin-weighted spherical harmonics can be found in 
Ref.~\cite{Beneke:2010eg}. 
With the solution of the homogeneous equation at hand, we can immediately write
down the solution of the inhomogeneous equation:
\begin{eqnarray}
\nonumber
\Delta_{X,lm}(\eta)
&=&
\int d\eta'\,
\sum_{l_1, l_2} \sum_{X' = I,E,B}
e^{-\kappa(\eta, \eta')} \,i^{l-l_1-l_2}\frac{(2l+1)(2l_1+1)}{2l_2+1}
j_{l_1}(k(\eta-\eta'))\\
&&\times\,\left(\begin{array}{ccc}l & l_1 &l_2 \\ m & 0 & m\end{array}\right)
\left(\begin{array}{ccc}l&l_1&l_2\\F_X&0&F_{X'}\end{array}\right)
H^*_{XX'}(l-l_1-l_2)\rho_{X',l_2 m}(\eta').
\end{eqnarray}
Using the multi-index notation from the main text, we can define the function
\begin{align}
\label{losgreen2}
\nonumber
j_{nm}(x) =&
\sum \limits_{l_1} i^{l_n-l_1-l_m} \frac{(2l_n+1)(2l_1+1)}{2l_m+1}j_{l_1}(x)
H_{X_n X_m}^*(l_n-l_1-l_m)\\
&\times
\left(\begin{array}{ccc}l_n & l_1 &l_m\\ m_n & 0 & m_m\end{array}\right)
\left(\begin{array}{ccc}l_n & l_1 &l_m\\ F_{X_n} & 0 & F_{X_m}\end{array}
\right)
\end{align}
and $\rho_{(v_\mathrm e, 1m)}\equiv 0$ to obtain the compact equation
\begin{equation}
\label{lossolution2}
\Delta_{n}(\eta_0)
= \int_0^{\eta_0}  d\eta \,e^{-\kappa(\eta)} 
j_{nm}(k(\eta_0-\eta))\rho_m(\eta),
\end{equation}
which holds for $X_n = I, E, B$.


\end{document}